\documentclass[]{aastex-jmc}
\usepackage{ifthen}
\usepackage{rotating}
\usepackage{lscape}
\usepackage[onecolumn]{epsfig,emulateapj5.jmc}
\usepackage{amsmath}		
\usepackage{timing}
\usepackage{ISM}

\newif\ifnotes
\notestrue
\notesfalse

\newif\iflong
\longtrue
\longfalse

\newif\ifshort				
\shorttrue
\iflong
  \shortfalse
\fi

\newcommand{\That}{$\rm T$}		
\newcommand{\Beff}{B_{\rm eff}}
\newcommand{\Wms}{W_{\rm ms}}
\newcommand{\Wims}{W_{i, \rm ms}}

\newcommand{\that}{\hat t}
\newcommand{\DMhat}{\widehat {\DM}}
\newcommand{\Deff}{D_{\rm eff}}

\newcommand{\nup}{\nu^{\prime}}
\newcommand{\Dp}{D^{\prime}}
\newcommand{\xpvec}{{\bf x^{\prime}}}

\newcommand{\XPBF}{X_{\rm PBF}}
\newcommand{\hatXPBF}{\hat X_{\rm PBF}}

\newcommand{\Ssys}{S_{\rm sys}}

\newcommand{\taudscaling}{\displaystyle\frac{2\beta}{\beta-2}}
\newcommand{\scalingA}{\displaystyle\frac{2}{\beta-2}}
\newcommand{\scalingB}{\displaystyle\frac{2\beta-1}{\beta-2}}
\newcommand{\scalingC}{\displaystyle\frac{\beta}{\beta-2}}
\newcommand{\scalingD}{\displaystyle\frac{\beta-1}{\beta-2}}

\newcommand{\Xarray}{\bf X}
\newcommand{\Carray}{\bf C}
\newcommand{\Dvec}{\bf D}

\begin{document}

\singlespace
\parindent 0pt
\newcommand{\be}{\begin{eqnarray}}
\newcommand{\ee}{\end{eqnarray}}
\newcommand{\etal}{et al.}

\title{
	A Measurement Model for Precision Pulsar Timing}
\author{J. M. Cordes and R. M. Shannon}
\shortauthors{Cordes \& Shannon}
\affil{Astronomy Department, Cornell University, Ithaca, NY 14853}
\email{cordes@astro.cornell.edu}
\email{ryans@astro.cornell.edu}

\begin{abstract}
This paper describes a comprehensive measurement model for 
the error budget of pulse  arrival times.
Applications include forecasting timing errors for optimization of
long-duration timing campaigns and identifying 
mitigation methods that can reduce the errors.
The model takes into account many  end-to-end effects
from the neutron star  to the telescope, 
with emphasis on  plasma propagation effects 
(particularly interstellar scattering),
which are stochastic in time and 
have diverse dependences on radio frequency.  
To reduce their
contribution, timing measurements can be
made over a range of frequencies that depends on a variety
of pulsar and instrumentation-dependent factors that we identify. 
A salient trend for high signal-to-noise measurements of
millisecond pulsars is that 
time-of-arrival precision
is limited either by irreducible interstellar scattering or by 
pulse-phase jitter caused by variable emission 
within pulsar magnetospheres.  Forecasts that ignore these 
contributions are likely to be overoptimistic.
A cap on timing errors implies that pulsars must be confined to 
low dispersion measures and observed at  high frequencies. 
Use of wider bandwidths that increase
signal-to-noise ratios  
will degrade timing precision, after a point, if nondispersive chromatic
effects are not mitigated. 
The allowable region in the dispersion measure-frequency plane 
depends on how chromatic timing perturbations 
are addressed in wideband measurements.   
Without mitigation, observations at 1.4~GHz or 5~GHz are restricted
to $\DM\lesssim 30$ and $\lesssim 100~\DMu$, respectively.   With aggressive
mitigation of interstellar scattering
and use of large telescopes to provide adequate sensitivity at
high frequencies (e.g. Arecibo, FAST, phase 1 of the SKA,
and the SKA), pulsars with DMs up to 500~$\DMu$ can be used in
precision timing applications.   
We analyze several methods that involve wide-band fitting to arrival
times at a given epoch prior to multi-epoch fitting. 
While the terms of greatest current astrophysical interest are achromatic
(e.g. orbital and gravitational wave perturbations),
their measurement
may ultimately be limited by similarly achromatic  stochasticity 
in a pulsar's spin rate. 
\end{abstract}

\newcommand{\donotes}{1}


\section{Introduction}
\label{sec:intro}

Timing of pulsars commenced with their original discovery over 40 years ago
and has reaped enormous benefits in enabling 
fundamental tests of general relativity
\cite[][]{2006Sci...314...97K}, 
demonstrating
the existence of gravitational waves
 \cite[][]{1982ApJ...253..908T}, 
determining masses and other fundamental aspects of neutron stars (NSs)
\citep[][]{2009arXiv0902.2891F}, 
and modeling of
the interstellar medium (ISM)
\citep{c+90, r+06}.   
The precision of time-of-arrival
(TOA) measurements has improved progressively, allowing new areas of
study to be targetted, such as higher-order relativistic corrections to orbits
and the timing perturbation from low-frequency (nano-Hertz) gravitational waves
\citep{d79, hd83, rt83}.
The discovery of millisecond pulsars (MSPs) 
\citep{b+82} opened the door to placing meaningful
limits on or making  detections of cosmological gravitational-wave (GW)
backgrounds \citep{bcr83, cs84, fb90, s+90, k+94}.
Projections based on improved timing methods and discovery of new
MSPs now makes the detection of nano-Hz GWs plausible in the next decade 
\citep{c+04b,j+04,k+04,j+05a,j+05b, l+08, j+09, h+09, v+09, h+10}.

This paper is motivated by the general goal of maximizing timing precision 
but especially for GW detection.
The sensitivity of
a pulsar-timing array (PTA) increases nonlinearly with data span length
but likely 
requires 5 to 10 years to make detections of plausible GW backgrounds
\citep{a+09, vH+09}. 
 The relevant GW frequencies
are therefore in the $10^{-8} - 10^{-9}$~Hz range.  Candidate signals include
stochastic backgrounds from coalescences of supermassive 
black-hole binaries \cite[][]{2003ApJ...583..616J, sv10} 
in galaxy mergers and from exotic sources, such as cosmic string
interactions \cite[][]{2005PhRvD..71f3510D,pt10}.   Precision timing
has also been proposed as a means for detecting low-mass
primordial black holes ($\lesssim 10^{-3}$ Earth masses) 
that make close flybys of the solar system \citep{sc07}.     
Individual GW  sources may be  detectable if they are
sufficiently massive and nearby \cite[][]{fl10}.   
No matter the source, the targeted signals
are very weak, requiring timing precisions along with stability of the 
pulsar clocks $\lesssim 100$~ns for a large number of MSPs over many years.

We present a measurement model that allows an end-to-end
assessment of timing precision.
We distinguish between phenomena that affect 
the time-tagging of pulses --- TOA estimation --- and 
the astrophysics of the pulsar clock.   
An illustration of this  distinction 
consists of pulses that can be time-tagged to arbitrary precision, 
such as delta-function pulses with 
infinite signal-to-noise ratio, but where  
stochasticity in the pulsar spin rate and in interstellar 
delays introduces a random component into the arrival times.
We thus  separate terms in the timing equation
that result from the quality of the astrophysical clock 
and over which we have no control (except by choice of pulsars) 
from those that can be mitigated
through choices of observational parameters and by
appropriate post processing.   These include 
phase jitter intrinsic to the pulsar, instrumental effects,
and  all the strongly chromatic
interstellar propagation effects.
We acknowledge but do not analyze timing errors from instrumental
polarization, time transfer, and solar system ephemerides, which are
discussed in the literature and likely will be improved to the point
of becoming secondary factors in timing precision.

Timing perturbations that are interstellar in origin are one of the
main topics of this paper.  \cite{a84} first pointed out
that stochastic dispersion-measure variations 
could limit timing precision.
A timing and interstellar scattering study \citep{c+90} of
the MSP B1937+21 (J1939+2134), 
indicated that 
diffractive interstellar scintilltions (DISS) cause
fast timing perturbations and that  
epoch dependent interstellar variations also occur.
Diffractive TOA variations were also identified in observations
and simulations by \cite{2010ApJ...717.1206C}.
\cite{fc90} assessed the roles
of dispersion measure and refraction-angle variations 
and concluded that non-dispersive interstellar effects need
to be mitigated in precision timing programs. 
This was followed by a similar study by \cite{1991ApJ...366L..33H}.
More recently,  \cite{y+07} demonstrated the 
importance of correcting for dispersion measure variations 
in precision pulsar timing.

Our analysis leads to discussion of possible methods for
removing most of the interstellar effects along with some of those
intrinsic to the pulsar that affect TOA estimation.   
If measurements were made with the same telescope and
the same intrumentation in a long timing program and that, 
hypothetically, the ISM caused chromatic perturbations that
were epoch independent, much of the mitigation effort 
would be unnecessary.
  However,  ISM effects {\em are}  inherently stochastic
and PTAs are expected to make use of multiple telescopes  with
instrumentation operating in different frequency
bands  that likely will change with time.   Consequently
mitigation of ISM effects must be done carefully. 

Our studies also apply to cases
where interstellar scattering limits the timing precision to 
milliseconds or worse, which is relevant to any pulsars in the
Galactic center even when observed at frequencies above 10~GHz.     

In a separate paper \citep[][hereafter Paper I]{sc10} we assess
the role of spin noise in precision pulsar timing, which we mention 
here only briefly.   A companion paper (Shannon \& Cordes 2010, in
preparation; hereafter Paper III) presents simulations
that corroborate and probe more deeply 
the interstellar plasma effects we discuss here analytically.  
We defer to another paper a consolidated analysis of the requirements
of a PTA for making detections of GWs. 

In \S~2 we summarize prominent physical contributions to arrival times.
In \S~3 we focus on achromatic and weakly chromatic timing perturbations. 
\S~4 concentrates on strongly chromatic effects.
In \S~5 we consolidate various physical terms into a signal model for use
in fitting arrival times. 
In \S~6 we discuss the efficacy of different schemes for correcting 
arrival times for 
chromatic interstellar contributions.
We summarize our conclusions in \S~7.
Appendices \ref{app:jitter} and \ref{app:screens}  
present details of several contributions to arrival-time
estimation errors.  
Appendix \ref{app:cone} considers the apparent frequency dependence
of dispersion measures that results from scattering. 
Appendix \ref{app:lsq} gives the details of
a least-squares fitting formalism. 

\section{Terms in a Physical Model for Pulse Phase}
\label{sec:terms}

Timing equations have been presented 
\citep[e.g.][]{bh86, 2006MNRAS.372.1549E}
that give a comprehensive inventory of deterministic contributions
to the overall arrival time and, indeed, motivate
timing studies in the first place.
However, they have only 
incompletely addressed the timing error
budget. Improvements in timing precision are always being strived for
that allow new effects to be measured. 
We describe salient terms in the timing budget that must 
be considered in any precision
timing analysis.  We first categorize them in terms of their
physical origin.  Then we consolidate terms according to their
statistical properties and wavelength dependence.  
Table~\ref{tab:timing_effects} lists
a comprehensive (but by no means exhaustive) set of 
terms and defines notation for deterministic and stochastic
components.
We designate the chromaticity of each effect, the
spectral signature, and whether there is a
correlation between different pulsar lines of sight.
Entries in Table~\ref{tab:timing_effects}  
are labelled as
astrophysical (`A' in column 2)  if they determine the arrival 
time. Others are designated with `\That'  
if they contribute to TOA estimation errors.
Some effects of course fall in both categories.  
Here we describe briefly some of the systematic 
astrophysical factors 
that determine the mean arrival time.  A detailed discussion 
of interstellar perturbations is given in later sections.

Pulse phase is dominated by the spin rate of the NS 
and its deceleration by magnetospheric torques.
The stochastic part includes 
glitches and timing noise, the latter typically
showing low-freqency components with a red power spectrum and more rarely
a bandpass spectrum \citep{h+09, h+10}.   Timing noise may comprise torque 
pulses or step functions that arise from effects internal 
to the neutron star (superfluid-crust interaction) or within 
the magnetosphere. 
Recently \cite{l+10} have shown that torque fluctuations in the
form of jumps between discrete states at quasi-periodic intervals
comprise much (but not all) of the timing noise in long-period pulsars. 
Spin fluctuations clearly are a nuisance for the use of pulsars
as precision clocks, as is well known.   
The power spectrum for spin variations in 
some pulsars is degenerate or nearly so with that expected from
some sources of cosmological GW backgrounds \citep{sc10}, so unless
timing noise can be removed, it will play a significant role in
the choice of pulsars for PTAs.  \cite{l+10} show that distinct pulse
shapes are correlated with the discrete torque states in some objects.
The correlation must be close to unity to remove most of the timing noise
and so far it is not clear that the correlation is large enough for
timing noise to be ignorable in forecasts of timing precision. 

Pulses emitted from the rotating
magnetosphere are proxies for spin phase of the NS
(with due allowance for any doppler shift from NS motion),
but emission regions in the magnetosphere are not synchronized
perfectly  with the NS.
Arrival times are (usually) calculated 
by fitting a template to pulse profiles 
obtained by averaging $N$ pulses.
Individual pulses are well
known to vary in amplitude and pulse phase by large amounts, 
but by mechanisms that are not well understood.  However they 
likely involve incomplete filling of the open-field
line region in the magnetosphere by coherently radiating plasma
\citep[][]{rs75}.
Radio emission may occur at different altitudes in response to 
time variable pair plasma densities, which will change arrival times
in accord with light-travel times, angular aberration, gravitational
bending of ray paths, and any refraction in the magnetosphere
\citep{ba86}.
Arrival time perturbations from the finite number of pulses
combined with intrinsic phase jitter imply that there is a correlation
of TOA error with pulse shape.  These changes in shape are much
different in origin from those identified by \cite{l+10} and signify
changes in pulse phase rather than torque.  Nonetheless, a sufficiently
high correlation again implies another opportunity for correcting 
TOAs, which we explore elsewhere.  

The evolution of pulse shapes with frequency includes changes
in widths, separations and amplitudes of individual pulse components
that comprise average profiles. 
Pulse widths typically scale with frequency as
$W\propto \nu^{-x}$ with $x\lesssim 0.3$
\cite[][]{2002ApJ...577..322M}, so observations that strive for
TOA precisions of 1\% of the pulse width or better require
mitigation of pulse shape changes.
A key issue is the alignment of frequency-dependent profiles so that
TOAs are consistent between frequencies.
One prescription is to identify fiducial pulse phases at different
frequencies  so that the resulting
TOAs match those expected from the cold-plasma dispersion law,
$\propto \nu^{-2}$  \citep{c70}.   
This procedure will fail at the
highest timing precisions because there are  other frequency-dependent,
interstellar time delays that scale differently with frequency    
\citep{rnb86, cpl86, fc90}. 
A recent discussion of the interaction between pulse shape evolution 
with frequency and estimates of dispersion measures may be found
in \cite{amg07}.   
While the focus of this paper is on frequency-dependent plasma effects,
a complete analysis must include {\em all} pulse shape changes.    

Other entries in Table~\ref{tab:timing_effects} correspond to
deterministic contributions for orbital motion, 
gravitational lensing, and astrometric errors associated with
referencing TOAs to the solar system barycenter (SSBC).  
Any error in the location of the SSBC from uncertainties in
planetary perturbations or from passing sub-stellar objects
\citep[e.g.][]{sc07} will produce a perturbation 
having a dipolar signature, 
which we designate as $\dtNewtSSBC$.
Stochastic perturbations from any orbital 
debris (R. Shannon et al., in preparation)
 and instrumental effects from time transfer and
polarization calibration errors \citep[e.g.][]{b2000, vS09} are included
in the table.

While all listed phenomena must be confronted in precision
timing programs, this paper focuses on those that affect estimation
of individual TOAs, including pulse shape stability, 
additive noise, and especially delays that arise
from dispersion, scattering and refraction in the interstellar plasma.


\begin{landscape}
\begin{deluxetable}{lclclcccccl}
\tabletypesize{\footnotesize}
\tablewidth{700pt}
\tablecaption{\label{tab:timing_effects}
Selected Timing Effects}
\tablecolumns{11}
\tablehead{
\colhead{Term} 
& \colhead{Type$^a$} 
& \multicolumn{2}{c}{Mean Part} 
& \multicolumn{2}{c}{Stochastic Part} 
& \colhead{Achromatic} 
& \multicolumn{2}{c}{Fluctuation Spectrum} 
& \colhead{PSR-PSR}   
\\
&&\multicolumn{2}{c}{---------------------------------}
 &\multicolumn{2}{c}{---------------------------------}
& \colhead{or} 
& \multicolumn{2}{c}{---------------------------------}	
& \colhead{Correlation$^{d}$} 
& \colhead{Comments}
\\
& & \colhead{Symbol} 
& \colhead{Value}
& \colhead{Symbol} 
& \colhead{Value}
& \colhead{Chromatic$^{b}$} 
& \colhead{Signature$^{c}$}
& \colhead{Shape}
}
\startdata
Spin rate     & A &
	 $\tspin$ &yr& $\dtspin$ &$\mu s - s$& a &   B, R & $f^{-4}-f^{-6}$ & U & \\
Magnetosphere: \\ 
\quad\quad Pulse Shape  & A, \That &
	 $\tP$ & $\rm\mu s - ms$& ---         &---& c  &  --- &---   & U & $\nu^{-0.3}$\\
\quad\quad Pulse Jitter & A, \That &
	---        &---& $\dtJ$    &$<\rm\mu s - ms$& c & W, B &see text& U & $\nu^{-0.3}$ \\
Orbital	  & A &
	$\torb$  &hr& $\dtorb$  &$\rm < ms$& a &   L, R & $f^{-5/3}$ & U & \\
Dispersion  & A, \That &
	$\tDM$   &$\lesssim s$& $\dtDM$   &$\rm\lesssim 100 \mu s$& 
	C  &   R & $f^{-5/3}$ & U & $\nu^{-2}$ \\
Faraday Rotation & A, \That & $\tRM$   &$\rm \lesssim \mu s$& $\dtRM$   
	&$\lesssim ns$& 
	C  &   R & $f^{-5/3}$ & U & $\nu^{-3}$ \\
Interstellar Turbulence & 
	\\
\quad\quad
   Pulse Broadening  & A, \That &
	 $\tPBF$ &$ns - s$& $\dtPBF$  &$\rm <ns - ms$& C & --- & complex & U & $\nu^{-4.4}$\\
\quad\quad
   DISS & A, \That & ---
	 	&---& $\dtPBFDISS$ &$\rm\lesssim \mu s$& 
		C & W  &flat& U & $\nu^{-1.6}$ - $\nu^{-4.4}$ \\
\quad\quad
   RISS & A, \That & $\tPBFRISS$ &$\rm\lesssim\mu s$& $\dtPBFRISS$ &$\lesssim \mu s$
		& C & R & $f^{-7/3}$ & U & ? \\
\quad\quad
   Angle of Arrival & A, \That &
	--- & --- & $\dtAOA$ &$\rm\lesssim \mu s$& C 
	& R &$f^{-2/3}$ & U & $\nu^{-4}$\\
\quad\quad
   Angle of Arrival & A, \That &
	---  & --- & $\dtAOABary$ &$\rm\lesssim \mu s$& C 
	& R &$f^{-1/3}$ & U & $\nu^{-2}$\\
\quad\quad
   Multipath averaging & A, \That &
	--- & --- & $\dtDMnu$ & $\rm \lesssim 0.1\mu s$ & C
	& R & complex & U & $\nu^{-23/6}$ \\
Astrometric$^e$ & \That &
	$\tAST$ &&$\dtAST$  && a &   --- & --- & U & \\
Newtonian solar perturbations & \That &
	--- & --- & $\dtNewtSSBC$ && a & --- & --- & C & dipolar \\
Radiometer Noise & \That &
	 ---         &---& $\dtRN$ &$\rm<\mu s - ms$& c$\to$C & W & flat& U &  
	$\nu^0\to\nu^{-2.7}$ \\ 
Polarization & \That &
	 ---          && $\dtpol$&& c & W & flat & U & \\
Gravitational
   Lensing   & A &
	 $\tGL$ && $\dtGL$  && a & --- & --- & U & Episodic\\
Cosmic Strings & A & $\tSTR$ && ---& & a & R & $f^{-16/3}$ & U & 
	Red noise if \\ 
	&&&&&&&&&& multiple events \\
\hline \\
Gravitational
   Waves     & A &
	   ---       &---& $\dtGW$  &$\rm\lesssim 100~ns$& a 
		&  R & $f^{-13/3}$& C, U & Two terms\\
\enddata
\tablenotetext{a}{A = astrophysical, \That = timing estimation error}
\tablenotetext{b}{a = achromatic, C = strongly chromatic, c = weakly chromatic}
\tablenotetext{c}{Fluctuation spectrum properties: R = red, W = white, B = bandpass, L = lowpass}
\tablenotetext{d}{U = uncorrelated between different pulsar lines of sight,
	C = correlated}
\tablenotetext{e}{Includes clock errors and Earth spin variations}	
\end{deluxetable}
\end{landscape}
    
\section{Achromatic and Weakly Chromatic TOA Perturbations}
\label{sec:achromatic}

The consolidated stochastic and achromatic term (including weakly chromatic contributions listed in Table~\ref{tab:timing_effects}) is 
\be
\Dt_{S,A} = 
	  \dtJ
	+ \dtAST
	+ \dtRN
	+ \dtpol.
\ee

In the following we are most interested in terms that appear even when
astrometric contributions and instrumental polarization
are dealt with perfectly. 

Assuming that astrometric and polarimetric errors are negligible
after appropriate processing,
a simplified stochastic, achromatic term that includes
only pulse phase  jitter and radiometer noise, is
\be
\Dt_{S,A} = \dtJ + \dtRN.
\ee
Both terms affect the shape of the pulse profile formed from
a finite number of pulses and thus contribute to the random
error in the arrival time.

\subsection{Radiometer Noise}
Radiometer noise adds 
to the pulse shape measured in any of the Stokes
parameters.  The resulting 
TOA error standard template fitting 
\citep{dr83} is approximately $W/\SNR$, 
where $W$ is the pulse width and $\SNR$ is the
signal to noise ratio of the pulse.  The frequency dependence of
the sky background is strong ($T_{\rm sky} \propto \nu^{-2.7}$) at
frequencies below about 0.5~GHz.  However, for higher frequencies usually
used for precision timing, the frequency dependence is weaker.
The signal-to-noise ratio 
involves the pulsar's radio spectrum that typically also declines with
frequency, canceling much of the variation of the sky brightness
temperature with frequency.  
We designate radiometer noise as weakly chromatic 
in Table~\ref{tab:timing_effects} 
and bundle it with stochastic achromatic terms.
In some cases, however,  the pulsar's spectrum
is flat \citep{lylg95},  
so the signal-to-noise ratio can be strongly chromatic.

In Appendix~\ref{app:rn} we derive 
the {\em minimum} TOA  error
that applies when the
conditions for matched filtering are satisfied, namely that the measured
pulse shape comprises an invariant templae shape  added to noise; 
our analysis assumes that
the noise is white.  
For a sum of $N = 10^6 N_6$ pulses,  
is
\be
\dtRN = 
 1~ \mu s \Wms N_6^{-1/2}\SNR_1^{-1} (\Delta/W)^{1/2} 
=
0.71~ \mu s~ 	\left(\nu/1.4~{\rm GHz}\right)^{-\alpha_p} 
		P_{\rm ms}^{-1} \Wms^{3/2} (B\,N_6)^{-1/2}
        	\left(S_{\rm sys}/ S_{1400} \right).
\label{eq:RN}
\ee
In the first form, 
$\SNR_1$ is the signal-to-noise ratio of a single pulse
(defined as the ratio of pulse peak amplitude to rms noise) and
$W$ is the effective pulse width defined in Appendix~\ref{app:rn},  
expressed here with a fiducial value of 1~ms. 
For a Gaussian pulse shape the effective width is 
$W = W_{\rm FWHM}/\sqrt{2\pi\ln2} \approx  0.69 W_{\rm FWHM}$.  
The quantity $\Delta$ is approximately
equal to the time resolution.
In the second form, which shows explicitly that $\dtRN$ is independent
of $\Delta$,
$S_{\rm sys}$ is the system temperature expressed in Jy;
$S_{1400}$ and $\alpha_p$  are the period-averaged flux density in mJy and  
the spectral index, respectively; 
the radio frequency  $\nu$ and bandwidth $B$ are in GHz; 
and the spin period $P$ is in ms. 

The appropriate values for pulse width $W$ and $\SNR_1$ 
include any distortions that broaden the pulse and reduce its amplitude.
Residual dispersion smearing is one such effect, although it is negligible
if pulses are  coherently dedispersed.  Broadening from 
interstellar scattering of distant pulsars observed at low frequencies
makes the pulse asymmetric and thus non-Gaussian, 
so the expression is only approximately correct. 

Another way that the ISM affects the timing error is through 
DISS of low-DM pulsars observed at high frequencies, which can cause
the apparent flux density to be much smaller or much larger
than average.    TOAs are sometimes obtained only during 
scintillation maxima
\citep[e.g.][]{hbo06}
which can enhance the 
SNR by two orders of magnitude, reducing 
the noise-induced error below other contributions.

\subsection{Pulse Jitter and Amplitude Modulations in Known Pulsars}
\label{sec:jitterknown}

All well-studied canonical pulsars (CPs) show phase jitter and amplitude
modulations like that 
in Figure~4 of \cite{m+02}.   
Individual pulse amplitudes vary by 100\% (or more)
and pulse phases jump by of order the single-pulse width. 
Empirically, the average pulse shape of a pulsar 
is as much determined by the PDF of the phase jitter as it is by
the shapes of individual pulses, a description first made by
\cite{c70}.  
The commonly-used intensity modulation index $m_I$
(the rms intensity divided by the mean as a function of pulse phase),
is typically of order unity and quantifies both the amplitude
modulation and the phase jitter.
Phase jitter appears to be statistically independent
between pulses in some objects, while
others show systematic drifts over several to many pulse periods, 
yielding an effective number of independent pulses $N_i \le N$ 
in an $N$-pulse average. 
We distinguish jitter from sustained ``mode''  changes in pulse shape
displayed by some pulsars, where pulse shapes 
switch between two or three preferred shapes.   
We quantify
phase jitter alone; if mode changes are unaccounted for in TOA estimates,
they will only exacerbate the timing estimation error.
We use the dimensionless quantity $f_J$ to quantify the rms phase jitter
in terms of the intrinsic pulse width, $W_i$.

Objects other than CPs also show phase jitter. 
The Crab pulsar shows giant pulses
with amplitudes several hundred times larger than the mean pulse amplitude;
giant pulses are intrinically narrow ($1-10~\mu s$) but show TOA jitter
$\sim 100~\mu s$ \citep{l+95, s+99, c+04a}.   
Giant pulses from
the MSP B1937+21 show $f_J \approx 0.2-0.5$ \citep{kt00}.  
However, in stark contrast, the ordinary (non-giant) pulses from this pulsar 
apparently show negligible jitter \citep{jap01, jg04}.  Other MSPs show jitter
similar to that of CPs, including J0437-4715, the brightest known MSP
\citep{j+98}.  \cite{es03} show that 
the autocorrelation function (ACF) of the average profile 
for the recycled
41-ms pulsar J1518+4904 
is about twice the width of  the ACF of single pulses (their figure 9), 
an effect that
can be accounted for only  with phase jitter.

Jitter is related to pulse-shape stability that has been quantified
using the cross correlation coefficient $\rho(N)$ between a template 
profile and sub-averages
of $N$ pulses \citep{hmt75, rr95}.  These analyses typically
show $1-\rho$ declining as $N^{-1}$ 
for some ranges of $N$ 
(as expected for statistically
independent phase jitter between pulses) 
while in some objects the trend is flatter
for small $N$, signifying correlations between pulses.  
It can be shown
that the correlation coefficient is
\be
\rho(N=1) = \left[(1+m_I^2)(W_U/ W_a) \right]^{1/2}
	= \left(1-f_J^2\right)^{1/4} / \left(1 + m_I^2 \right)^{1/2},
\ee
where
$W_U$ is the characteristic width of the template profile, 
$W_a$ is the characteristic width of an individual pulse  
and we have used the relation, $f_J^2 = 1 - W_a^2/W_U^2$.
Typically with $m_a = 1$ and $W_U = 2W_a$, we expect
$\rho(N=1) = 1/2$.   For large $N$ we have 
$\rho(N) \approx 1 - (1/2N)(1+m_a^2)(W_U/ W_a)$.  
Correlation values reported in previous stability analyses are consistent
with these estimates, indicating that
$f_J\sim 1/3$~to~1/2 for most pulsars.  

\subsection{Timing Error from Pulse-phase Jitter}
\label{sec:jittermain}

When TOAs are calculated from averages of N pulses, 
any given profile will depart stochastically from a long-term average,
even when the SNR is large.   Template fitting
therefore yields a jitter-induced error that adds quadratically
 to the minimum
expected from matched filtering.
The role of jitter in pulse-shape stability and timing has been discussed 
in the past \citep{cd85, da+93, c93}, but little overall attention 
has been given to its role
in precision timing.  

The TOA error $\propto W_i/\sqrt{N}$ 
is pulsar dependent.
(Appendix~\ref{app:jitter}). 
Because single pulses from many pulsars appear to be broadband with
modest frequency dependence, 
the jitter timing perturbation in a typical data set is only
slightly dependent on frequency and can be viewed as a stochastic element
of the systematic profile evolution in frequency described earlier.   
For simplicity, in some  of our analysis 
we will consider the jitter effect to be stochastic in time but
constant in frequency.   
For Gaussian-shaped pulses and
a Gaussian probability density function (PDF) for pulse phase jitter,
\be
\dtJ =
0.28\mu s\,  \Wims N_6^{-1/2} 
		\left(\frac{f_J}{1/3}\right)
		\left(\frac{{1+m_I^2}}{2}\right)^{1/2}. 
\label{eq:jitter1}
\ee

\begin{figure}[h!]
\begin{center}
   \includegraphics[scale=0.40, angle=0]{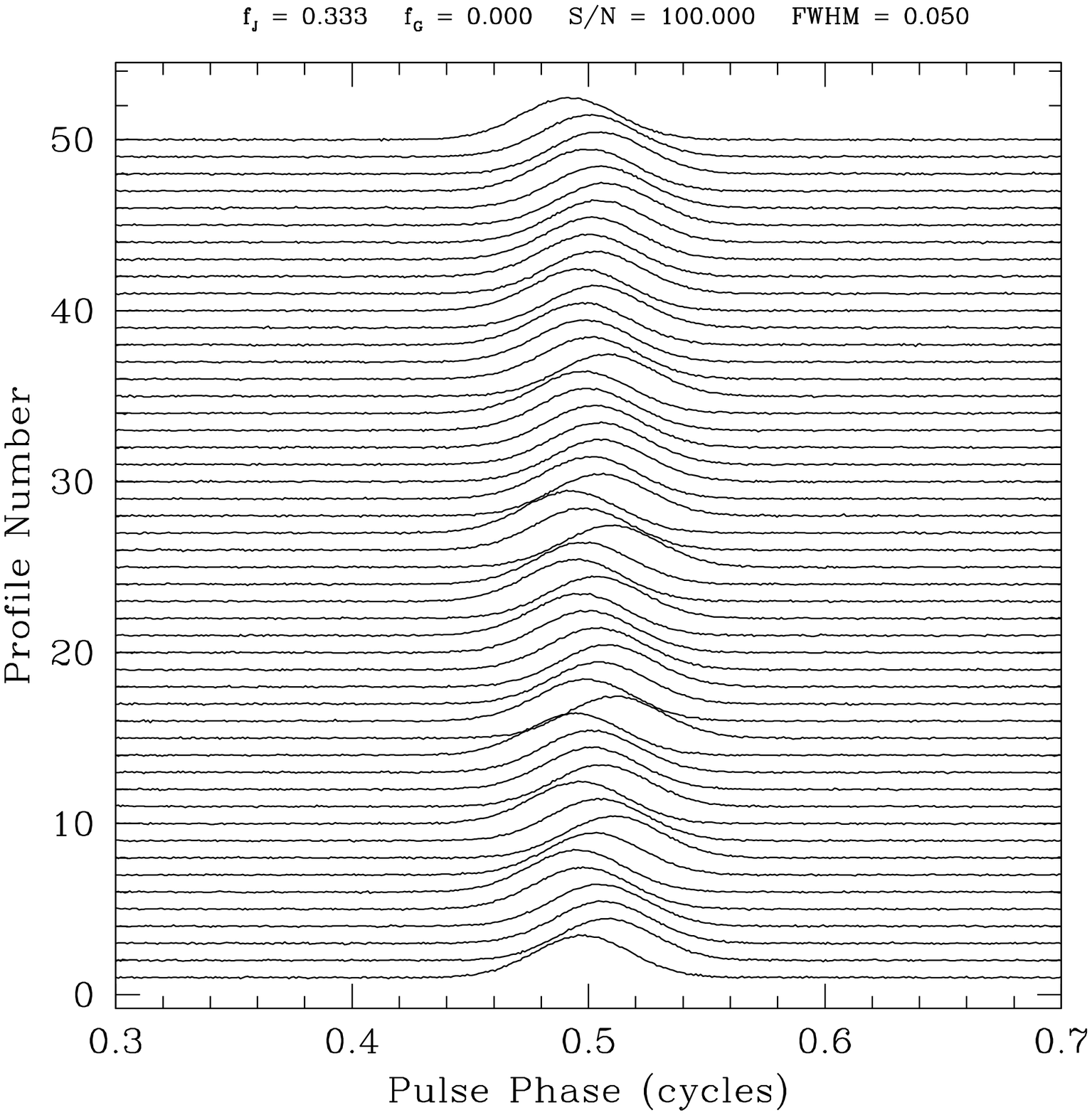}
   \includegraphics[scale=0.40, angle=0]{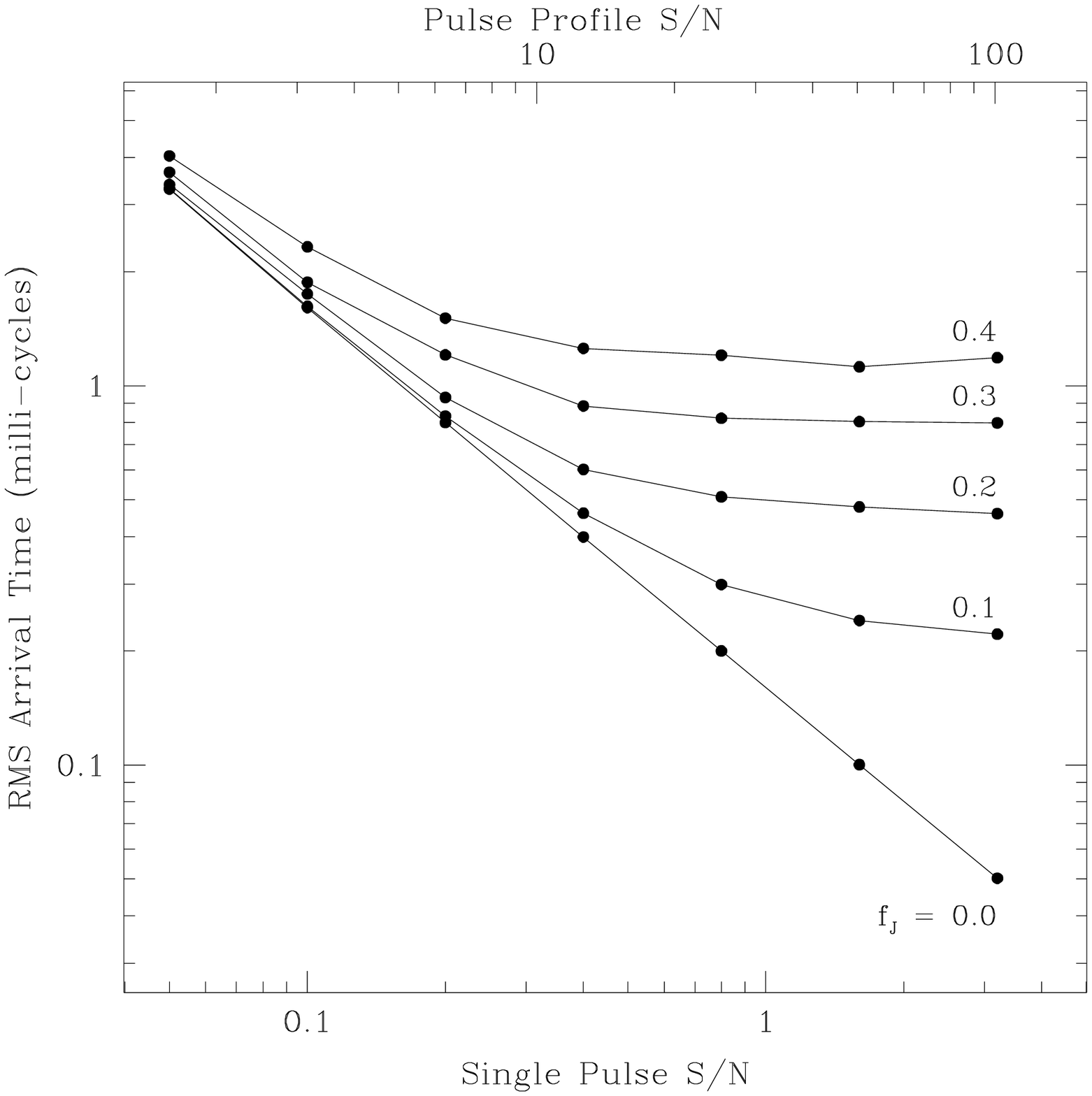}
   \caption{
	(Left) Simulated pulse sequence of showing
		pulse phase jitter with $f_J = 1/3$ and a unity
		intensity modulation index.   The net duty cycle
		is 0.05.
	\label{fig:jitter-sequence}
	(Right) Arrival time error vs signal to noise ratio (SNR)  for
		simulated data with different amounts of pulse phase jitter,
		as labelled.   TOAs were calculated from averages
		of $10^3$ pulses to yield  a SNR for each profile
		that is $10^{3/2}$ larger than the single-pulse SNR.
		Each plotted point is based on 500 realizations.
	\label{fig:rms_toa_vs_snr}
   }
\end{center}
\end{figure}

Figure~\ref{fig:rms_toa_vs_snr} shows the rms TOA error caused by different
levels of phase jitter from simulations.
The left-hand panel shows a sequence of pulses that displays pulse-phase
jitter with $f_J = 1/3$.  The right-hand panel shows the rms
arrival-time error for pulse profiles formed from a fixed number
of single pulses ($10^3$) as a function of signal to noise ratio 
and for different values of $f_J$. For $f_J = 1/3$,
the rms error is double that from additive noise once the 
single-pulse SNR exceeds a few tenths, which is consistent with 
a comparison of Eq.~\ref{eq:RN}, \ref{eq:jitter1}.

\begin{figure}[h!]
\begin{center}
   \includegraphics[scale=0.40, angle=0]{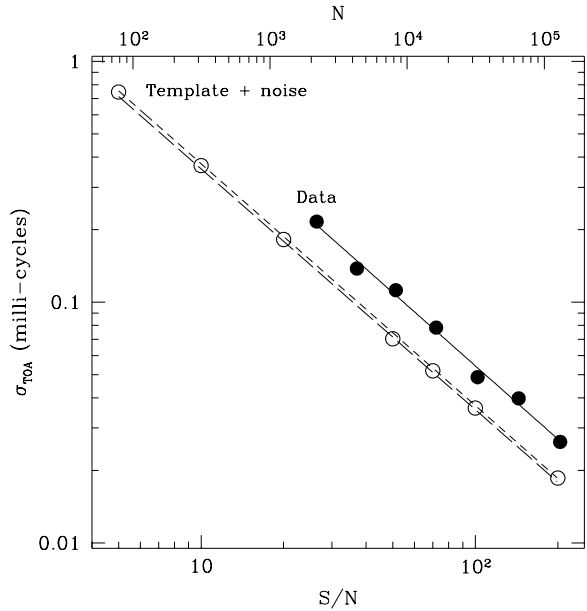}
   \caption{
 	Arrival time analysis for the MSP J1713+0747
	at 1.4~GHz.
	The points are calculated for pulse profiles calculated from
	different numbers of pulses ($N$, top axis) yielding
	increasing signal-to-noise ratios (bottom axis).
	The lower set of points results from calculating TOAs from
	simulated data formed by adding a 
	high SNR template to gaussian noise. 
	The upper set of points
	is from fits to actual data obtained with different integration
	times.   
	The short-dashed curve is the fit to the points. 
	The long-dashed curve
	is the prediction using Eq.~\ref{eq:dtRN}-\ref{eq:Weff}, 
	which assumes that
	data comprise identical pulse shapes with additive noise.
	The two curves are actually coincident but have been displaced
	by $\pm 0.01$ in the log for clarity.   
	The actual data have higher TOA errors than
	simulated values because pulse-phase jitter causes
	pulse shapes to differ from the template shape.   
	A jitter parameter $f_J\approx 0.4$ accounts for the difference.   
	\label{fig:J1713+0747}
   }
\end{center}
\end{figure}

In Figure~\ref{fig:J1713+0747} we show a jitter analysis for
the MSP J1713+0747. 
Data were obtained at the Arecibo Observatory at 1.38~GHz using
a baseband recording system, coherently dedispersed to yield 
high-time-resolution,  and averaged synchronously using a timing
model from P. Demorest and D. Nice 
\citep[][private communication]{2007PhDT........14D}.  
Details of the data acquisition and other analyses will
be described elsewhere (Shannon et al., in preparation). 
We compare TOA errors expected from radiometer noise alone to those
actually measured. First we take a high S/N template profile, add
white noise, and then calculate the arrival time.  Repeating over
a large number of realizations and for different amounts of noise
yields the open circles shown in Figure~\ref{fig:J1713+0747}.   
We use Eq.~\ref{eq:RN} and calculate the effective width using
the template profile and Eq.~\ref{eq:Weff}.  The numerical
rms TOA error agrees very well with the predicted error.   
Actual data show excess TOA variations.   We calculated TOAs for
an initial set of 10-s averages (2188 pulse periods) 
and then for 20-s, 40-s averages, etc.   
A jitter parameter, estimated assuming $m_I = 1$,
\be
f_J = 
     	\left[N \left(\sigma_{\rm TOA}^2(N)-\sigma_{\rm TF}(N)^2\right) /
		  \left(1+m_{\rm I}^2\right)
	\right]^{1/2} 
	\approx  0.4 
\ee
accounts for the excess timing error over that expected from template
fitting (TF) to a profile consisting of the template combined with
additive noise to yield the same SNR as the real data.  

\subsection{Jitter vs Noise Dominated Timing Errors}

\begin{figure}[h!]
\begin{center}
   \includegraphics[scale=0.40, angle=0]{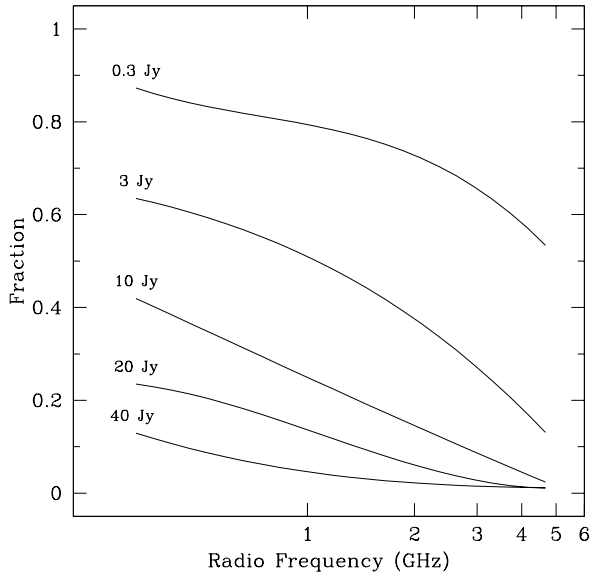}
   \caption{Fraction of known pulsars with jitter timing error 
	larger than that from radiometer noise, i.e.
	$\dtJ > \dtRN$.   The curves are labeled by the system
	equivalent flux density, $S_{\rm sys}$,  at 1.4~GHz.    
	$S_{\rm sys}$ varies with frequency by adding the Galactic
	synchrotron contribution as $20~{\rm K}~(\nu/0.408~{\rm GHz})^{-2.75}$
	as quantified by \cite{c07}.    
	Calculations were done with 10\% bandwidths below 1~GHz and
	bandwidths of 0.4 and 0.8 GHz at 1.4 and 2~GHz, and 1~GHz bandwidths
	at 3~GHz and higher.
	Each MSP with $P< 20$~ms that had  
	tabulated flux densities  in
	the pulsar catalog of \cite{m+05} 
	is included.   Given that only 82 MSPs in the current
	catalog satisfy the criteria, a cubic fit was done to the 
	computed fraction to plot each curve. 
	\label{fig:fraction}
	}
\end{center}
\end{figure}

We compare jitter and noise contributions to the TOA error 
for the known sample of pulsars by using cataloged parameters \citep{m+05} 
and assuming reasonable values for amplitude modulation 
($m_{\rm I} = 1$) and phase 
jitter ($f_J = 1/3$). 
Figure~\ref{fig:fraction}
gives the fraction of MSPs for which
phase jitter dominates the radiometer noise error.   We assume
that jitter is statistically independent between pulses so
that $N_i = N$, which may {\em underestimate} the TOA error.
Fractions are
calculated by counting objects for which $\dtJ >\dtRN$  
for different system temperatures (in flux-density units, Janskys) 
that correspond to 100-m class telescopes (e.g. the Effelsberg
and Green Bank Telescopes, the Expanded VLA, and the Parkes and Sardinia
telescopes),
Arecibo-class telescopes (Arecibo itself,
the Chinese FAST telescope, and Phase 1 of the Square
Kilometer Array, SKA), and Phase 2 of the SKA, which 
is projected to have $S_{\rm sys}$ of order ten times
smaller than for Phase 1. 
Results are shown for frequencies from 0.3 to 5 GHz  using either
the cataloged spectral index of the pulsar's flux density  or using a power law
$\propto \nu^{-2.0}$.  

The results indicate that usage of more sensitive telescopes
(whether by larger collecting area, lower system noise, or greater 
bandwidths) will  lead to timing measurements that are limited
by phase jitter.   To increase the timing precision,
the only recourse is 
to make use of longer integration times per TOA.
It may also be possible to reduce the timing error from phase jitter
if the resulting pulse shape variations are correlated with
the timing error.   A method based on principal component analysis
will be reported separately (Cordes et al., in preparation). 

\section{Strongly Chromatic TOA Perturbations}
\label{sec:chromatic}

Stochastic and strongly chromatic timing perturbations include 
contributions from plasma dispersion, Faraday rotation and 
interstellar scattering (ISS),
\be
\Dt_{S,C} = 
	  \dtDM + \dtRM + \dtISS.
\ee
The dispersion term receives contributions from the ionosphere,
interplanetary medium (IPM), and the ISM.  Any extragalactic pulsars
will also involve the intergalactic medium and the ISM of their
host galaxy.
The ISS term comprises six separate terms 
\be
\dtISS =  
	 \dtPBF 
	+ \dtAOA
	+ \dtAOABary
	+ \dtPBFDISS
	+ \dtPBFRISS
	+ \dtDMnu.
\label{eq:dtiss}
\ee
These are respectively 
the delay associated with diffractive pulse broadening, 
the delay from angle-of-arrival (AOA) variations, 
the error in referencing TOAs to the SSBC due to AOA variations,
the error from  fast variations in the pulse-broadening function (PBF) that
describes diffractive pulse broadening, slow variations in the
PBF induced by refraction, and the perturbation of DM that results from
averaging over the scattering cone.  
We discuss these terms in  detail in \S~\ref{sec:ISS}.


The relative amplitudes of ISS effects depend on the spatial properties
of the ISM.   The standard approach is to characterize small-scale
variations in electron density 
($\lesssim 1000$~AU) 
--- those that diffract
and refract radiation from point sources --- with a wavenumber spectrum
and to consider large-scale variations 
(e.g. cloud-like, sheet-like and filamentary structures) 
as modulations of the spectrum. 
Small scale variations largely appear to
be consistent with a Kolmogorov spectrum and in many cases the variations 
involve anisotropic irregularities  with large axial ratios
\cite[][]{2006ApJ...637..346C,b+10}.
Regardless of the detailed shape of the wavenumber
spectrum, so long as it is broad, three regimes of scattering can 
be defined \citep{r90, cl91}.   When the phase perturbation by the scattering
medium is much smaller than a radian across one Fresnel
scale $\sim \sqrt{\lambda D_{\rm eff}}$, where $\lambda$ is the wavelength
and $D_{\rm eff}$ is a characteristic distance, a point-source will 
appear largely as a point source with a small halo around it.    
When the phase perturbation is large, the scattering is strong.
All of the source flux will be scattered and 
the scintillation bandwidth  $\dnud\ll\nu$.    So long as the scattering
length $\ell_d = \lambda/\theta_d$ is larger than any minimum (inner) 
scale $\ell_i$ in the wavenumber spectrum, the scaling of $\dnud$ with $\nu$ is
a function of the spectrum's slope (if shallow enough).   For 
super-strong scattering, $\ell_d < \ell_1$ and the scaling laws generally
will be different than in the strong-scattering regime. 
In particular, they will conform to a medium with a square-law 
phase structure function.
Electron density
fluctuations appear to exist on a wide range of scales from $\sim 10^3$~km
to Galactic scales so, accordingly, there is a large range of expected
time scales \citep{ars95}.  


When scattering is weak, only a fraction of the pulsar's flux
is scattered, the remainder appearing as a point source.  
In this case, the delay of the pulse is negligible because even
the scattered  portion is delayed by only $(2\pi\nu)^{-1} \approx 0.1$~ns. 
The transition from weak to strong scattering occurs approximately
when 
$\dnud/\nu = 1$ or 
$\nu_{\rm trans} = C_1 / (2\pi\taudbar)$.  
Using $\taud = {\taud}_0 \nu^{-4.4}$, the broadening time
at 1~GHz and with $\nu$ in GHz,
(the Kolmogorov scaling when there is neglibile inner scale), we
obtain 
$\nu_{\rm trans} = \left(2\pi{\taud}_0 / C_1 \right)^{0.29}
= 12.1~{\rm GHz}~ {\taud}_0(\mu s)^{0.29}$.   
From Figure~\ref{fig:tvsDM}, 
which shows values of $\taud$ at 1~GHz plotted against
DM,   the transition frequency is about 2.3, 12 and 24~GHz for pulsars
with $\DM$s of 10, 50 and 100~$\DMu$.

\subsection{Dispersion Measure Variations}
\label{sec:dDM}


For a total \DM\ 
from all contributions along the line of sight (LOS),
the group delay is
\be
\tDM = a_{\rm DM}~ \DM~ \nu^{-2} 
\label{eq:tDM}
\ee
where $a_{\rm DM} = 4.15~{\rm ms}$
for \DM\ in standard units of pc~cm$^{-3}$.
Fluctuations $\delta\DM(t)$ affect the arrival time itself and, 
if large enough and unaccounted for, can distort the pulse shape
and affect the TOA. 
In the following we ignore such distortion because it is addressible in
the data processing.
\DM\ variations produced by
structure in the ISM  are stochastic in time but 
deterministic in frequency (however, see \S~\ref{sec:screenave}).  
Measured variations are seen on a variety of time scales that are consistent
with a broad spectrum of electron-density variations in the ISM
\citep{a84,c+90, r90, pw91, b+93, k+94, cl97}. 
Amplitudes depend on the distance and velocity of the pulsar but
maximum rates of change $\sim 10^{-3}~ \DMu$ per year.
The temporal fluctuation spectrum for DM has the same form as 
the electron-density spectrum \citep{a84} and has large fluctuations
on longer time scales.   
During a typical observing session at a given epoch (e.g. $\lesssim 1$~hr),  
DM variations are too small to be detected with known pulsars. 
As is well known, however, precision timing requires removal
of DM variations between epochs \citep[e.g.][]{y+07}.

The precision needed for DM is stringent. The simplest
case is when the only other
TOA perturbation is from white-noise measurement error.  
We write the TOA as the sum of achromatic, dispersive,
and white-noise terms:
\be
t_{\nu} = t_{\infty} + t_{\DM}(\nu) + \tWHITE(\nu), 
\label{eq:tnu1}
\ee 
where $t_{\infty}$ is the achromatic arrival time at
infinite frequency (which would include spin noise and gravitational
wave contributions).  Here we ignore any frequency dependent
profile ``evolution'' intrinsic to the pulsar. 
For a two-parameter ($t_{\infty}$ and $\DM$) 
least-squares fit across a frequency
range $[\nu_2, \nu_1]$ sampled in $n_{\nu}$ channels,
the error on the achromatic TOA is
$\sigma_{t_{\infty}}^2 = s_4 /(s_0 s_4-s_2^2)$, 
where $s_p$ is a weighted sum over frequencies defined in Eq.~\ref{eq:sp}.
For equal errors in all channels 
$\sigma_{t_{\infty}}$ depends only on the frequency ratio 
$r=\nu_1/\nu_2 > 1$, as expected.  
An alternative method would use a value for DM obtained 
from ancillary, non-contemporaneous measurements
to correct TOAs rather than fitting for DM from the TOAs.
For this
procedure to be at least as good as fitting for DM,
the error in $\delta \DM$ must satisfy
\be
\delta\DM < 10^{-3.38}\,\DMu\,
	\frac{\nu_1^3}{B}(1-B/\nu_1)^{3/2}
	\sigma_{B}(\mu s)
\label{eq:dDM}
\ee
for $\nu_1$ and $B = \nu_1-\nu_2$ in GHz and a fiducial radiometer
noise TOA error $\sigma_{B} = 1~\mu s$ (c.f. Eq.~\ref{eq:RN}).  
In this expression, 
we assume that errors are independent of frequency
and emphasize that this is just the radiometer noise contribution to
the error.

Measured DM changes are at levels expected
for interstellar electron-density variations that have a power-law wavenumber 
spectrum.  For a power-law spectrum $\cnsq \times ({\rm wavenumber})^{-\beta}$,
where  $\cnsq$ is simply a spectral coefficient in the wavenumber spectrum,
 the structure function for DM  has an ensemble average, denoted by angular
brackets \citep[e.g.][]{cr98, y+07},  
\be
D_{\DM}(\tau) = 
	\left\langle
	 	\left[DM(t) - DM(t+\tau)  \right]^2
	\right\rangle 
	= 
f_{\alpha} \int_0^D ds\, 
	\vert \veffperpvec(s) \vert^{\alpha} \cnsq(s)
\approx
\frac{f_{\alpha}}{\alpha+1} \SM \left(\tau V_p\right)^{\alpha},	
\ee
where $\alpha = \beta - 2$, $f_{\alpha}$ is a constant, 
$\veffperpvec$ is an effective velocity, 
and $\SM$, the scattering measure,  is the LOS integral of $\cnsq$.  
The approximate equality holds
for constant $\cnsq$ and when
the pulsar transverse velocity dominates the effective velocity
$\veffperp$.
For a Kolmogorov medium,  		
$f_{5/3} = 88.3$, 
and using standard units for scattering measure
$(\SMu)$, the rms difference in DM between
two epochs separated by $\tau$ (in years) is
\be
\sigma_{\DM}(\tau) = \left[(1/2) D_{\DM}(\tau) \right]^{1/2}
	\approx 10^{-3.47}~\DMu~ 
	\left(\SM/10^{-3.5}~\SMu \right)^{1/2}
	\left(\tau_{\rm yr}{V_p}_{100} \right)^{5/6}
\label{eq:sigmaDM}
\ee
for pulsar velocity
in units of 100~km~s$^{-1}$.  This rms is comparable to measured
changes in DM for nearby pulsars that have $\SM \approx 10^{-3.5}~\SMu$.  
From Eq.~\ref{eq:sigmaDM} and \ref{eq:dDM} if we require a timing error
less than 0.1~$\mu s$ for $3\sigma_{\DM}(\tau)$, then the 
DM and TOA measurements need to be made within
$\tau < 8$~days. 
This result is in accord with those determined numerically
in Paper III.



\subsection{Profile Splitting from Faraday Rotation}

The TOA perturbation from a magnetized
plasma includes a birefringent term 
that is equal and opposite for the two hands of circular polarization,
$\tRM = \pm 0.18~{\rm ns}~ \nu^{-3} \RM$
for \RM\ in standard units of rad~m$^{-2}$.
The empirical manifestation 
is {\em profile splitting} of the two polarizations.
For the extreme case of high-RM pulsars observed at low frequencies, the 
profile splitting produces a pseudo-circular 
polarization that would cause a sign-changing Stokes parameter 
$V = I_L - I_R$
even when the circular polarization intrinsic to the pulsar  is zero. 
If uncorrected, the total intensity profile
$I = I_L + I_R$ is broadened slightly but symmetrically relative
to the true TOA, 
so the TOA error  is not biased one way or another.  
Correction of this effect, if ever necessary, needs to be done
prior to summing the two hands of polarization and can be incorporated
into the coherent dedispersion process.
Values of \RM\ for Galactic lines of sight can range up to $10^3$~rad~m$^{-2}$
but for objects within 1~kpc, typical RMs are a few tens of
rad~m$^{-2}$  or less. 
The RM perturbation is therefore negligible compared to a 100-ns
timing error unless distant pulsars with large RMs are observed at
low frequencies.  

\subsection{Interstellar Scattering and Refraction }
\label{sec:ISS}

The interstellar timing perturbations in Eq.~\ref{eq:dtiss} 
result from refraction 
and multipath propagation caused by scattering from
small-scale electron-density variations.
The resulting timing perturbations have 
distinct time-frequency characteristics:
\begin{enumerate}
\itemsep -2pt
\item Variations in dispersion measure from frequency-dependent
	spatial averaging of the phase perturbation ($\dtDMnu$); 
\item The mean time delay associated with multipath broadening of pulses 
	from diffraction ($\dtPBF$);
\item Fast, stochastic changes in the diffractive
	delay associated with the finite number
	of ``scintles''
	in the sampled time-frequency plane ($\dtPBFDISS$);
\item A time delay associated with the stochastic
	refraction angle of arrival (AOA, $\dtAOA$);
\item An AOA induced error associated with transforming measured
TOAs to the SSBC ($\dtAOABary$), which arises from the difference between
the assumed and actual direction to the pulsar; 
and
\item Modulations of the mean time delay from wavefront curvature induced
	by large-scale variations, which also causes refractive
	interstellar scintillation (RISS; $\dtPBFRISS$).    
\end{enumerate}

The first effect is the modification of DM 
by the frequency-dependent spatial averaging of 
the electron-density along different propagation paths.
The second (typically dominant) effect is the delay introduced by 
the pulse broadening function
(PBF) that is convolved with the true pulse shape.
The PBF is causal, so the delay is always positive.  It is 
also strongly frequency dependent. 
The PBF is stochastic from refractive variations and from the 
scintillation fluctuations contained in the finite bandwidth
and integration time used to form a pulse profile.   The angle of
arrival (AOA) of the centroid of the bundle of multipath ``rays''  
varies from refraction in the ISM and adds another frequency-dependent
delay with a different frequency dependence.  Finally, the AOA induces an error
in the correction of topocentric TOAs to the SSBC \citep{fc90}.

We characterize the second and last three effects
as a mean delay that changes slowly
with time (hours to years).  
The third effect 
differs from the others because it involves
interstellar diffraction features that are statistically independent
between data sets used to calculate individual TOAs. 
We give a detailed analysis in the next section.



For a Kolmogorov medium in the strong-scattering regime,
refraction angles are much smaller than diffraction angles,
so the TOA perturbations
included in Eq.~\ref{eq:dtiss} may be ordered as
\be
	\dtDMnu
	\lesssim \dtAOA  
	\lesssim \dtAOABary
	\lesssim \dtPBFRISS
	\lesssim \dtPBFDISS
	\lesssim \dtPBF
\ee 
because the two smallest terms depend quadratically
and linearly on the refraction angle,
$\theta_r$,  while the other terms depend on powers of the 
diffraction angle, $\theta_d$.    
Scaling laws and estimates are given in \S~\ref{sec:ISS}.   
Alternative orderings of terms may apply for media
with enhanced refraction from non-Kolmogorov structures.

The scaling of the pulse broadening time with frequency
indeed shows departures for some objects from the scaling expected for a
Kolmogorov medium (L\"ohmer et al. 2003). 
These departures either
indicate actual departures from a Kolmogorov wavenumber spectrum 
or the breakdown of the assumption of an unbounded screen, which can 
give anomalous
scalings \cite[][]{cl01}.   \cite{lo+04} conclude that
anomalous scattering is associated with distant objects whereas
nearby objects generally show Kolmogorov  scalings (or nearly so).
One such scaling is for the  
scintillation bandwidth $\dnud$ with frequency \citep{cwb85}.
Another is the slope of the structure function for \DM\
\citep[e.g.][]{y+07}.




\subsection{Frequency-dependent DM Variations From Multipath Screen Averaging}
\label{sec:screenave}

Consider a simple screen model for the diffraction and refraction.
The diameter of the scattering cone at a phase
screen located a distance $\Dp = D - \ds$ from Earth for a pulsar distance $D$  
is $\ell_s \approx 2\Dp \theta_d$.  Measurement of DM toward a source
therefore involves spatial averaging over an area $\sim \ell_s^2$.  
This causes the measured DM to differ from that for the direct line
of sight through the screen.  Since $\theta_d \propto \nu^{-2.2}$
or so, this implies that the difference in DM is strongly frequency 
dependent.  By defining a weighting function
$W(\xvec, \nu)$ that is defined by the scattered image expressed with
a spatial argument (i.e. $\xvec = \Dp\thetavec$) and thus
has a width $\sim \ell_s$, it can be shown that
the rms difference in DM is a weighted integral over the 
phase structure function and  
\be
\delta\DM   
 \approx 10^{-1.05}~\DMu~ \SM\, {\Dp}^{5/6} \nu^{-11/6}
\ee 
for $\SM$ in standard units, $\Dp$ in kpc, and $\nu$ in GHz. 
The resulting TOA perturbation, which is stochastic with characteristic
time scale $\ell_s / \veffperp$ (the  same time scale as for  RISS),
has rms
\be
\dtDMnu
 \approx 0.12~\mu s~
	{\Dp}^{5/6} \nu^{-23/6}
	\left( \frac{\SM}{\SMu} \right).
\ee
While small, this effect results from the dispersion measure effectively
being a function of frequency. It is a steep, declining function
of  frequency  because the area of the scattering cone decreases rapidly
with frequency. The frequency dependence is nearly degenerate with
some of the other timing errors.

\subsection{ Pulse Broadening Function}
\label{sec:PBF}

Pulse broadening from multipath scattering in the ISM is described
by a pulse-broadening function (PBF) that is convolved with the
emitted pulse shape to produce the measured pulse. 

\begin{figure}[h!]
\begin{center}
   \includegraphics[scale=0.40, angle=0]{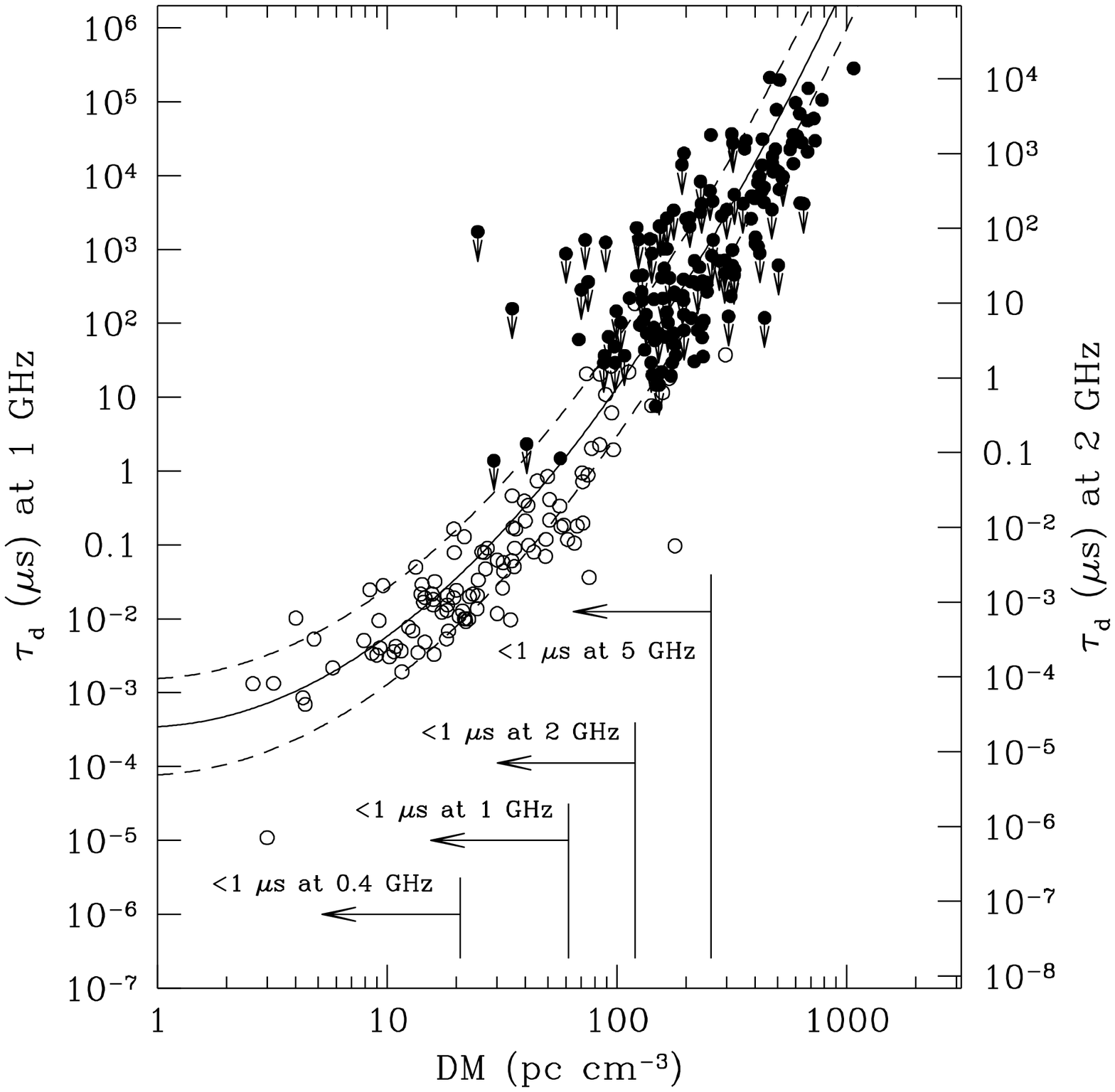}
   \includegraphics[scale=0.40, angle=0]{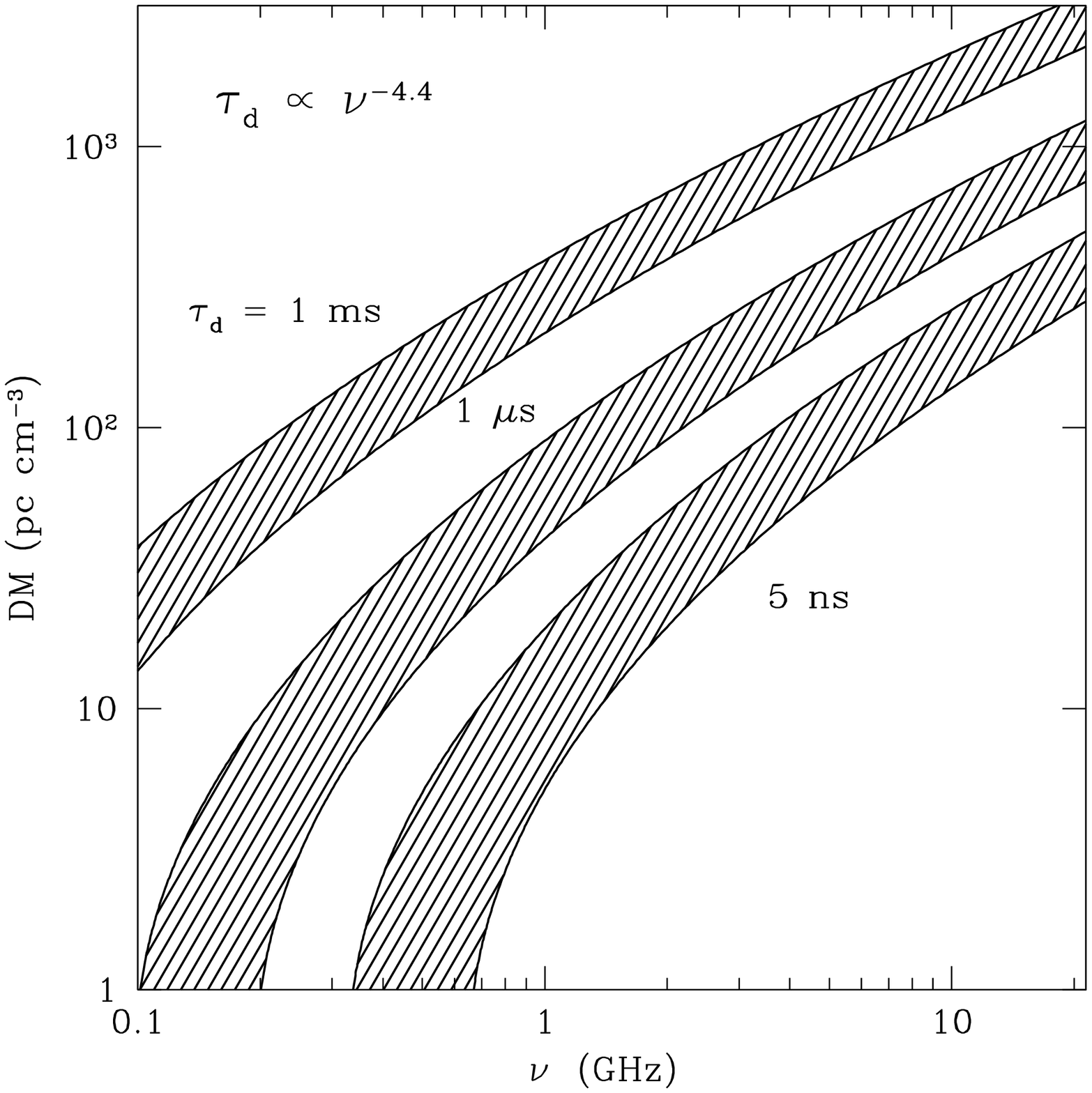}
   \caption{
	(Left)
	Pulse broadening time plotted against DM for pulsar data used to
	construct the NE2001 model \citep{cl02}.
	The solid curve shows the parabolic fit given in the text for
	$\log\taud$.
	and the dashed
	curves show $\pm 1.0\sigma$ deviations from the fit.  Open circles
	are inferred values using scintillation bandwidth measurements;
	filled circles are direct measurements of $\taud$.   
	Downward arrows denote upper limits. 
	All measurements have been scaled to 1~GHz assuming 
	$\tau_d\propto \nu^{-4.4}$ and the right-hand scale for 2~GHz 
	assumes the same scaling.  The strong scattering regime is
	assumed, which requires $\tau_d \gg C_1/2\pi\nu$ or
	$\tau_d \gg 2\times10^{-4}~\mu s$ at 1~GHz.  At lower frequencies,
	the entire range of $\tau_d$ in the plot satisfies the requirement.    
	Leftward arrows designate maximum values of DM for which $\taud < 1$~ms
	at 0.4, 1 and 2 GHz  using the nominal parabolic fit.
	\label{fig:tvsDM}
	(Right)
	Contours of constant pulse broadening time, $\tau_d$, using
	the fit shown in Figure~\ref{fig:tvsDM} including the
	$\pm 1\sigma$ uncertainty in the fit, which results in
	the widths of the shaded regions.  A $\taud\propto \nu^{-4.4}$
	scaling with frequency is used.  This scaling is steeper
	than appropriate for some high DM objects but is typical for
	low-DM pulsars.   
	\label{fig:regimes}
   }
\end{center}
\end{figure}

The ensemble average pulse-broadening function (PBF) has a characteristic
width $\taud \sim D\theta_d^2/2c$ where 
$\theta_d$ is the diffraction angle and $D$ is an effective distance.
The timing perturbation is proportional to the width of the PBF, 
which is strongly
wavelength dependent ($\propto \lambda^4$).  
The pulse broadening times 
in Figure~\ref{fig:tvsDM} 
demonstrate that they are a strong function of both DM and frequency. 
The empirical fit from \cite{b+04}is  
$\log\taud = -6.46 + 0.154\log\DM + 1.07(\log\DM)^2 - (3.86\pm 0.16)\log\nu$.
The scatter about this fit is $\pm 0.65$ in $\log\taud$.
We write the PBF as
$\PBF(t) = \PBFbar(t) + \delta\PBF(t)$.
where $\PBFbar$ is the ensemble-average 
shape\footnote{This might also be called a ``snapshot'' average
using the language of \cite{ng89}, which takes into
account that a statistically precise measurement at a given
epoch is still not a good ensemble average because there are
long-term refraction effects.  Our approach is to treat
refraction effects as a separate  modulation imposed on a statistically
precise average at a given epoch.}  
expected from a particular
medium and consider it to be a slow function of epoch.  By constrast,
$\delta\PBF$ encapsulates fast departures from the ensemble average.  
Defining $\PBF(t)$ to have unit area,
the characteristic broadening time is
\be
\taudbar =
        \int dt\, t\,\PBF(t) = \taud + \delta\taud,
\label{eq:taud}
\ee
and its ensemble average as $\taud$.

The scintillation bandwidth and the mean pulse broadening time $\taudbar$ are
related by\footnote{
Our definitions of $\dnud$ and $\taudbar$ follow 
\cite{cr98}, i.e. $\dnud$ is the HWHM of the intensity 
ACF vs. frequency lag and $\taudbar$ is the mean delay of
the PBF.  This differs from \cite{lr99}, who use alternative definitions
of $\taud$ in a similar expression and thus quote different 
values of $C_1$ than we use here.}
\be
2\pi\taudbar \dnud = C_1,
\label{eq:uncertainty}
\ee  
where $C_1$ depends on all the properties of the medium
\citep{cr98, lr00},
such as the wavenumber spectrum, its thickness and location along the
(LOS) and on its transverse extent \citep{cl01}.
For simple media, values of $C_1$ can be calculated.
For a {\em thin screen} unbounded transverse to the
LOS and with a circular Gaussian angular scattering function,
the PBF is a one-sided exponential function with $1/e$ time
scale $\taud$.  In this case the mean time delay is also $\taud$
so that $C_1 = 1$.
For media that are plausibly 
relevant to pulsar scattering, $C_1$ can vary by a factor of nearly two
\citep{cr98, lr99}.

\subsection{The Mean-Shift Regime}
\label{sec:shiftapprox}

A case of interest for nearby pulsars or observations
at high frequencies is where the pulse broadening
time is small compared to the pulse width, $\taud\ll W$
but where the
strong scintillation regime applies.  This requires 
that the scintillation bandwidth $\dnud\ll \nu$ or, equivalently,
$\taud\nu \gg 1$. 
We specify the mean-shift regime as
\be
\frac{C_1}{2\pi\nu} \ll \taud \ll W \quad\quad {\rm or} \quad\quad 
0.2~{\rm ns}~\nu_{\rm GHz}^{-1} \ll \taud \ll 1~{\rm ms}~W_{\rm ms}.
\ee 
Under these conditions the TOA perturbation is simply the pulse
broadening time calculated from Eq.~\ref{eq:taud} using the
instantaneous PBF (not the ensemble-average PBF).
The TOA perturbation also depends on 
the number of scintles
contributing to an observation.
In many cases it is large and
the TOA perturbation converges 
to some mean value.  This requires
that $\dnud \ll B$, where $B$ is the bandwidth, or that
$\dtISS \ll T$, where $T$ is the total time of the observation. 
Figure~\ref{fig:tvsDM} yields the range of DM for
which the shift approximation applies:
$\taud < 1$~ms for $\DM \lesssim 100$, 250 and 400~$\DMu$	
at 0.4, 1 and 2 GHz, respectively. 

One approach to timing precision is to require that
$\tau_d$ be smaller than some specified
rms timing error, $\sigma_{\rm max}$, which might be 100-ns or less
for pulsar timing array applications but could be much larger
for other timing purposes (such as pulsars in the Galactic center).
If $\tau_d < \sigma_{\rm max}$, pulse broadening might be ignored
entirely.   Alternatively, it might be corrected to a  
fractional precision $\epsilon_{\tau}$,  yielding a residual error
$\epsilon_{\tau}\tau_d< \sigma_{\rm max}$.  Using the empirical
fit for $\tau_d$ vs. DM shown in Figure~\ref{fig:tvsDM}, we can then
solve for $\DM(\nu)$ for different $\sigma_{\rm max}$.   Results
are shown in Figure~\ref{fig:regimes}.   
The smallest pulse broadening effects occur
in the lower-right portion of the diagram whereas
objects with large DMs are strongly affected by pulse broadening.
Many pulsars have steep spectra, so the choice of frequencies 
for a given object requires a compromise, as discussed in
\S~\ref{sec:mitigate}.

When the shift approximation holds, the TOA can be corrected
by subtracting $\taudbar$ from nominal values \citep{hs08}.
However, because by definition $\taudbar$ is small, 
we cannot measure it directly at the frequency of the TOA measurements.
There are several ways to estimate it, however.

\subsubsection{Correlation Approach}
\label{sec:corr}
The timing delay
$\taud$ can be estimated 
by using Eq.~\ref{eq:uncertainty}, where
the characteristic scintillation bandwidth
$\dnud$ is calculated as the half-width at half maximum of
the intensity correlation function $\Gamma_{\delta I}(\delta\nu)$
that is calculated from the dynamic spectrum $I(\nu, t)$.
This approach requires knowledge of the shape of the PBF 
so the type of medium must be known 
to determine the appropriate  value for $C_1$.
Alternatively, 
a characteristic time $\tausec$ can be calculated from the
secondary spectrum, the squared-magnitude of the Fourier transform of
the dynamic spectrum \citep[e.g.][]{s+01}, 
which is related to the autocorrelation function of the PBF.
The relationship of $\tausec$ to $\taud$ is also medium dependent
and requires knowledge of their ratio that is analogous to knowing $C_1$. 
The correction of TOAs for the mean PBF delay
is therefore problematic unless ancillary constraints on the scattering
physics 
are known, namely whether a thin screen or extended-medium
applies, whether the scattering region
is bounded transverse to the LOS, what the wavenumber spectrum is, 
etc.     
The secondary spectrum provides much of this information \citep{s+01, hs08}.

We identify the errors associated with this correction approach by
considering Eq.~\ref{eq:tnu1} with an additional
chromatic term, $t_C$, added.
For the case where $t_C$ consists solely of 
 the pulse broadening term and the mean-shift regime applies,
the TOA perturbation equals the pulse broadening time $\taud$.
The broadening time can be estimated from the scintillation bandwidth
obtained at a frequency $\nup$,
\be
\widehat {\taud} (\nup) = \frac{\hat C_1}{2\pi \widehat{\dnud}(\nup)},
\ee
where quantities requiring an estimate or an assumed value are hatted.
The broadening time can be scaled to other frequencies using
\be
\widehat{\taud}(\nu) = \widehat{\taud}(\nup)  
	\left (\frac{\nu}{\nup} \right)^{\displaystyle -{\hatXPBF}}.
\label{eq:tc1}
\ee 
The estimation error of the chromatic term and thus
on $t_{\infty}$ expands into terms
dependent on individual errors, 
$\delta C_1, \delta\dnud,$ and $\delta \XPBF$: 
\be
\delta \taud = \taud - \widehat {\taud} \approx 
	\taud
	\left[
	   \frac{\delta C_1}{C_1} 
	 - \frac{\delta\dnud}{\dnud}
	 - \delta \XPBF \ln \frac{\nu}{\nup}
	\right].
\ee
Based on estimates of scintillation bandwidths in the literature,
the range of possible values of $C_1$, 
and the error in the exponent, $\delta X_{\rm PBF}$,
the combined error in $\delta\taud$ is unlikely to be less than
10\% with this approach.
However, a 10\% error in removing pulse broadening implies that
the shaded bands in Figure~\ref{fig:regimes} will shift to the
left by about a factor of 1.8 in frequency (0.26 in $\log\nu$).

\subsubsection{Phase Retrieval Methods}

A second, empirical  approach requires much less
{\em a priori} knowledge about the scattering medium.  
It calculates the complete
PBF using only a measurement of the ACF of the PBF or, equivalently,
the magnitude of the PBF's Fourier transform.  
The ACF of the PBF can be obtained from the secondary spectrum, for 
example.  \cite{w+08} apply two-dimensional phase retrieval 
to the secondary spectrum 
and demonstrate the general principle.  
For the one-dimensional ACF with $n_{\tau}$ lags, 
there are $n_{\tau}$ functions that can produce the measured ACF.   
Many of these are acausal or
include negative values and can therefore be ruled out.
Those that remain include the function with minimum delay and its time
reverse, which has maximum delay.   
Methods exist to calculate the
{\em minimum-delay} PBF from the ACF, which is the unique member in
the large family of PBFs consistent with the ACF that has its amplitude 
most concentrated toward the origin  in a mean-square sense 
\citep{sca81}.
While physically a minimum-delay solution is not demanded by 
scattering and refracting geometries, it may be the most probable 
result given that
scattering and refraction angles have distributions peaked at zero.
Also it may provide a more accurate correction compared to use of 
an idealized form for the PBF used to provide a value for $C_1$ in
the correlation method.  
Application of inversion techniques will
be explored in a separate paper (Cordes \& Shannon, in preparation). 
Another approach (P. Demorest \& M. Walker, private communication)
investigates the phase of the wavefield directly to determine 
the PBF.



\subsection{Real World PBFs}

Time delays from pulse broadening presented so far represent
ensemble-average results. TOAs, however, are determined from
pulse profiles obtained as integrations over the time-frequency
plane corresponding to the integration time $T$ and bandwidth $B$. 
The $T-B$ plane contains a finite number of independent fluctuations
(``scintles'')
of the diffractive interstellar scintillations (DISS),
causing the instantanous PBF to differ from the ensemble average
shape, producing a statistical error in the estimated TOA,
as first demonstrated by \citep{c+90}.
In addition, because individual TOAs are typically estimated
from data sets with no overlap in $T-B$, the TOA error is statistically
independent and thus has a white-noise spectrum.

The TOA error arises from the summation of independent
scintles in the $T-B$ plane, the number of which can be large or small
when a pulse profile is obtained.
In the following we will use $\dtPBFDISS$ to mean both the specific
value applicable to a particular TOA and also to signify its rms value,
the context making it clear as which applies.
The rms PBF error is approximately 
\be
\dtPBFDISS = \Niss^{-1/2} \taud,
\label{eq:dtPBFDISS}
\ee
where $\Niss$ is the effective number of ``scintles.''  
Previous work estimates $\Niss$ as
\be
\Niss = \Nt\Nnu = 
	\left (1 + \eta_t T / \Dtiss \right)
	\left (1 + \eta_{\nu} B / \dnud \right),
\ee
where $\Dtiss$ is the characteristic DISS time scale and $\dnud$ is
the characteristic bandwidth, as before.  The filling factors
$\eta_t, \eta_{\nu}$ are less than unity and are in the range
of 0.1 to 0.3, depending on the definitions of the characteristic time scale
and bandwidth.  In the following we use approximate values for $\eta_t$
and $\eta_{\nu}$ while keeping in mind that  the factorization
of $\Niss$ is only an approximation.   In another paper 
(Cordes \& Shannon, in preparation), we derive exact
results for several cases  that validate the form of Eq.~\ref{eq:dtPBFDISS}. 

The scaling of the
timing variation, $\dtPBFDISS$, with frequency   
is different for the four cases where
$T$ and/or $B$ are larger or smaller than the characteristic scintle
size, i.e. whether  
$\Nt$ and/or $\Nnu$ are unity or much larger than unity and
whether the scattering is strong
or super-strong as defined in \S \ref{sec:ISS}.   In strong scattering
$\taud \propto \nu^{-\zeta}$ where
$\zeta = {2\beta/(\beta-2)}$ for $\beta<4$.  
From Eq.~\ref{eq:uncertainty}, the DISS bandwidth therefore
scales as $\nu^{\zeta}$. 
For a Kolmogorov medium ($\beta = 11/3$), $\zeta = 22/5$.  
In super-strong scattering,
the coherence length of the diffraction pattern
is smaller than the inner scale and $\zeta\to 4$.  
Similarly,
the characteristic DISS time scale scales as 
$\nu^{\kappa}$ with $\kappa = \beta / (\beta-2)$  in strong scattering 
and $\kappa = 2$ in super-strong scattering. 
The scaling  depends on the type of medium ($\beta$),
whether the scattering is strong or super-strong, and on the 
bandwidth and integration time relative to the DISS bandwidth and
time scale.  For a Kolmogorov medium
the index $x$ in $\dtPBFDISS \propto \nu^{-x}$
ranges between $22/5$ for
low-DM pulsars observed at relatively high frequencies (but where
the scattering is still strong) and $3/2$ for high-DM pulsars
observed at low frequencies in the super-strong scattering regime.
This corresponds to the case where the scattered wavefield's
coherence length is smaller than the inner scale of the wavenumber spectrum.   
In more detail  we have the following cases:
$N_t\ll 1$ and $N_{\nu}\ll 1$:  $x = 22/5$; 
$N_t\gg 1$ and $N_{\nu}\ll 1$:  $x = 19/5$; 
$N_t\ll 1$ and $N_{\nu}\gg 1$:  $x = 11/5$; 
$N_t\gg 1$ and $N_{\nu}\gg 1$ and negligible inner scale:  $x = 8/5$; 
$N_t\ll 1$ and $N_{\nu}\ll 1$ and where the diffraction scale
is smaller than the inner scale:  $x = 3/2$. 

\begin{figure}
\begin{center}
   \includegraphics[scale=0.40, angle=0]{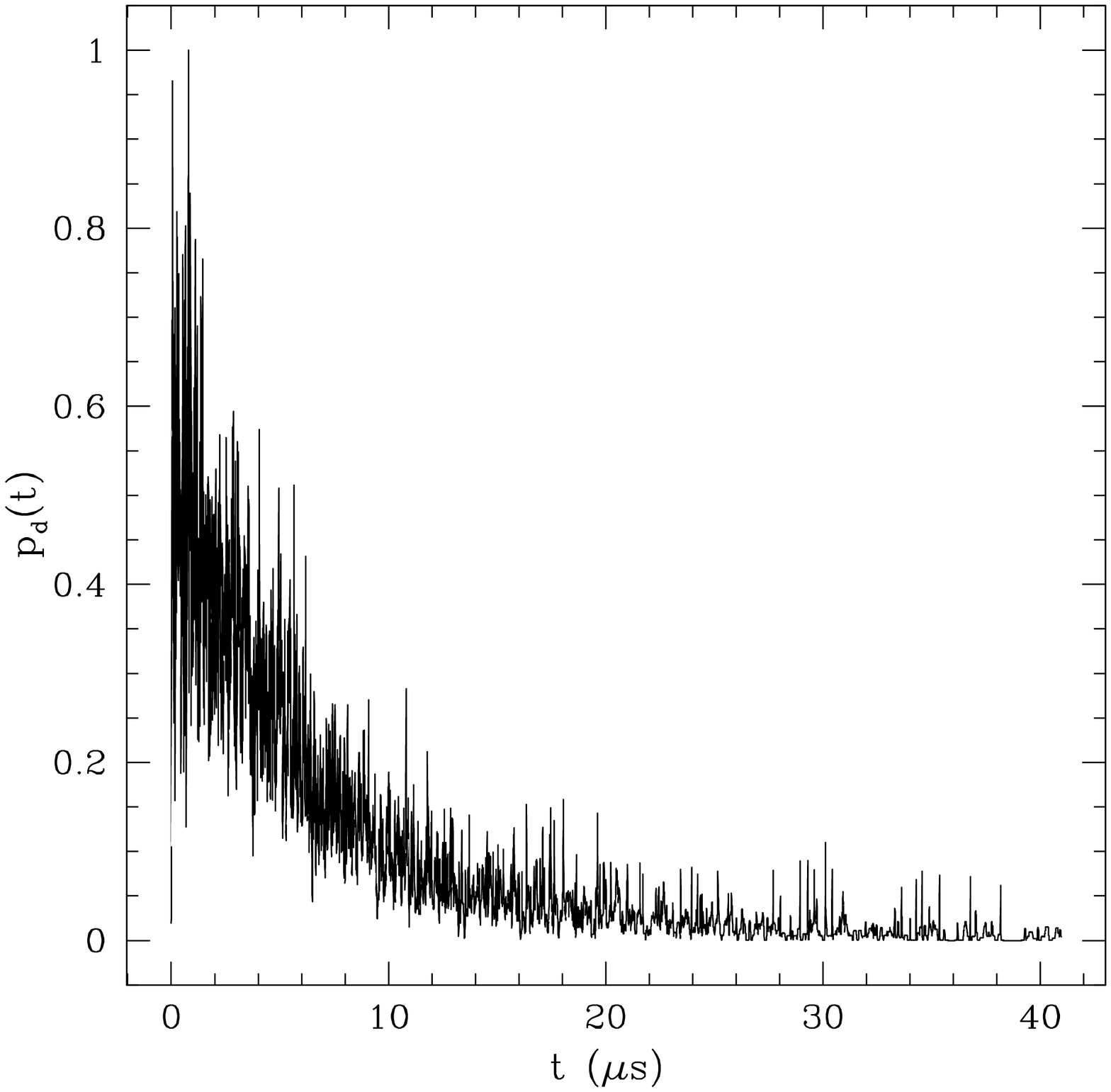}
   \includegraphics[scale=0.40, angle=0]{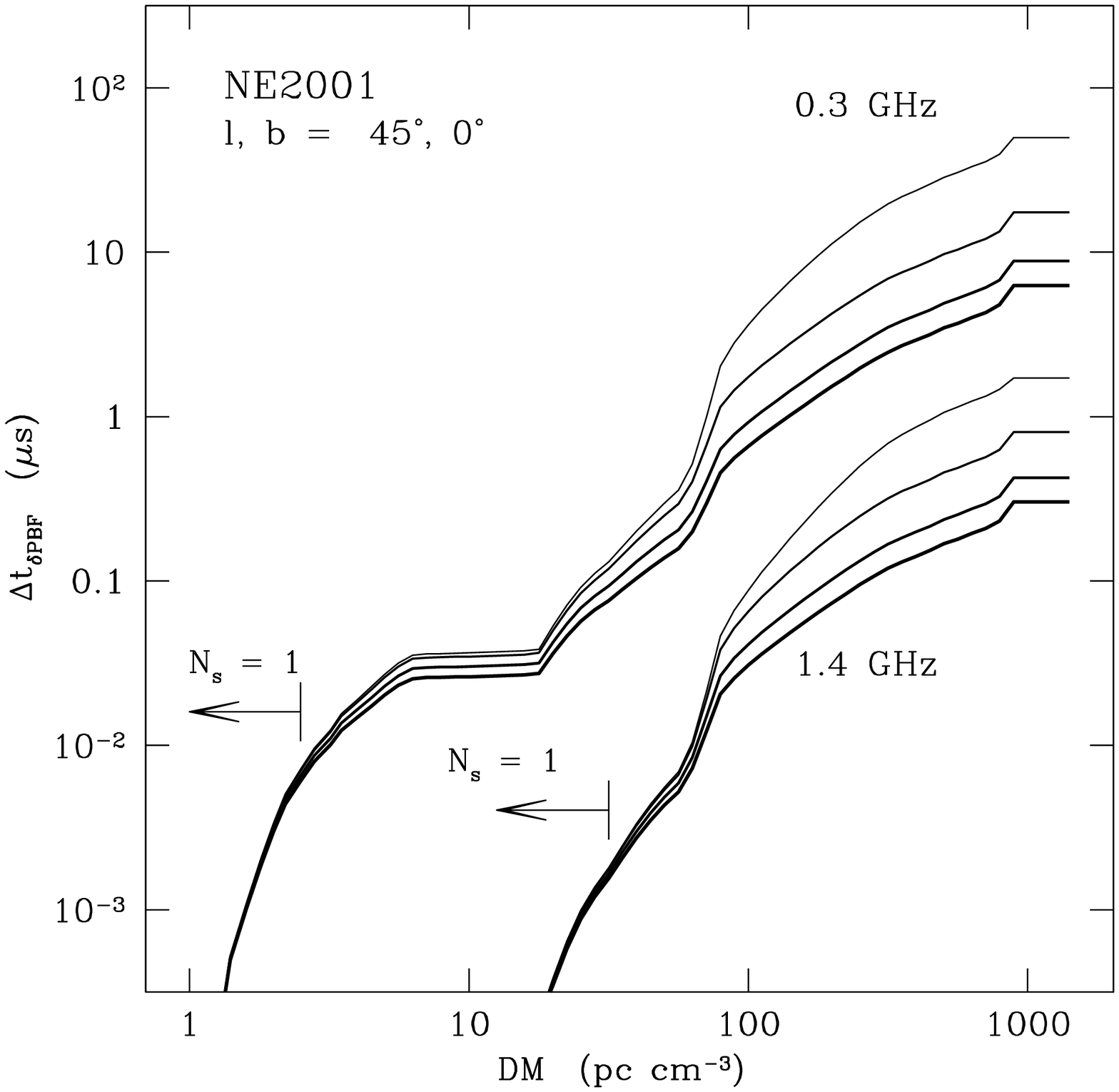}
   \caption{
	Left: Simulated pulse broadening function showing 
	stochastic variations that result from the finite number of
	scintles in the time-frequency plane.   This case is for
	a pulsar scattered into an rms angle of 1.25 mas by a screen
	1~kpc from both the pulsar and the observer, an observation
	frequency and bandwidth of 1.4 and 0.1~GHz, respectively,
	and a total observing time of 2~hr. Details of the simulation
	are available on request. 
	Right: Timing error due to variations in the pulse broadening
	function  from the finite number of scintles.  Results are
	shown for a LOS through the inner Galaxy 
	($\ell = 45^{\circ},~b = 0^{\circ}$) for two frequencies,
	0.3~GHz and 1.4~GHz, with bandwidths of 50~MHz and 200~MHz, 
	respectively.
	From top to bottom in each set, the curves are for effective velocities
	of 10, 100, 400 and 800~km~s$^{-1}$.  For lower velocities,
	there are fewer scintles averaged over the fixed observing time
	of $T=300$~s so the timing error is larger.  The arrows indicate
	values of \DM\ for which the number of scintles sampled
	$N_s = 1$.
	\label{fig:rmspbf}
   }
\end{center}
\end{figure}

Figure~\ref{fig:rmspbf} shows predicted values of 
$\dtPBFDISS$ as a function of DM 
for Galactic coordinates,
$\ell = 45^{\circ}, b = 0^{\circ}$, 
and particular observational parameters.  The NE2001 model for
the Galactic electron density \citep{cl02} was
used to calculate $\dtISS$ and $\dnud$ vs. DM.
Different curves correspond to different effective velocities 
that underly the scintillation time scale. 
The figure shows that the  finite-scintle TOA perturbation increases rapidly 
for more distant pulsars and for observations
at lower frequencies.  The explanation is that
even though the number of scintles increases
rapidly, the pulse broadening time $\tau_d$ increases even faster.

\subsection{Large Pulse Broadening Times}

For distant pulsars observed at low frequencies
and especially those with very narrow intrinsic pulses, the
pulse broadening time can exceed the pulse
width, causing strong distortion of    
th intrinsic pulse profile 
as well as delaying it. In this case, 
template fitting 
must be customized to each pulsar and frequency combination.
Scintles are too narrow in time and frequency
to resolve, so phase retrieval methods are not possible.
One approach is to deconvolve 
the PBF from the measured pulse profile \citep[e.g.][]{b+03} 
to yield the TOA along with determinations of the relevant PBF shape
and the intrinsic pulse shape.
This requires usage of a family of candidate PBFs for which the
deconvolution indicates a best fit that is non-negative.  The template bank of 
candidate PBFs needs to include  those that are truncated since
evidence exists, for highly scattered pulsars,
that scattering screens  are spatially truncated transverse to the LOS
\citep{lo+04}.

Consider a multifrequency timing program where the  
pulse broadening is larger than the pulse width $W$ 
at lower frequencies and smaller than $W$
at higher frequencies.  The purpose of such a program may be
to conduct precision timing by removing chromatic interstellar effects.
The lower frequency measurements could be used to identify the form
of the PBF, including  its characteristic delay, and the frequency scaling
of the delay.  Corrections of the TOAs for the PBF delays and 
for a contemporaneous dispersion measure DM(t) calculated from the
TOAs would then follow.  This case is discussed explicitly in
\S~\ref{sec:mitigate}.

Another case is where the pulse broadening is large at all frequencies
in the program.  This situation applies to pulsars in
the Galactic center because radio scattering is particularly intense
so that timing at radio frequencies $\lesssim 12$~GHz
will not be in the mean-shift regime.   TOA shifts from pulse
broadening will depend in detail on the templates used.  In addition,
orbital motion that changes the distance to the scattering material
that encloses the Galactic center will cause secular changes in
the pulse broadening.

\subsection{Angle-of-Arrival and Refractive Scintillation Variations}

Structures in the ISM refract radiation if they are much larger than the Fresnel scale 
$\sqrt{\lambda D} \approx 10^{11}$~cm and larger than
the so-called refractive or multipath scale $D_{\rm eff}\theta_d \approx~1$~AU, 
causing
the angle of arrival (AOA) to deviate from the direct path.
Wavefront curvature caused by the same large-scale structures
changes the flux density \citep{r90} to produce 
refractive interstellar scintillations (RISS).    
We give results here only for thin screen media, which serve 
as simple examples and allow scaling laws to be derived.


Let $\thetar$ be the departure of the AOA from the direct path
as viewed by an observer.  The
induced time delay for a fiducial refraction angle of 1~mas is
\be
\dtAOA &\approx& \frac{1}{2c}
	\Deff
	\thetar^2
	\approx 1.21~\mu s~ \Deff(\rm kpc) \thetar^2({\rm mas})
\ee
where $\Deff$ is the effective distance through the ISM. For
a thin screen at distance $D_s$ from a pulsar at  distance $D$
from the Earth,
$\Deff = \left ( D - D_s \right) \left(D/D_s\right)$.
(If the refraction angle is defined instead as the bending angle produced
by the screen, the effective distance is
$\Deff = (D-D_s)(D_s/D)$.)
The delay scales as $\lambda^4$ if refraction is produced by a discrete 
cloud but scales slightly differently for a medium with a large range
of scales \citep[e.g.][]{cpl86, rnb86}.


A second effect is associated with the transformation of the TOA to
the SSBC \citep{fc90}, which depends on an assumed
direction to the pulsar.  Interstellar refraction causes the actual
and true directions to differ, inducing a perturbation that
scales as $\lambda^2$,  
\be
\dtAOABary = c^{-1} \hat n \cdot {\mathbf r}_{\oplus}
	\approx  c^{-1}r_{\oplus} \thetar(t) \cos \Phi(t) 
	\approx 2.4~\mu s\, \thetar(\rm mas) \cos\Phi(t),
\ee
where $R_{\oplus}/c \approx 500$~s and $\Phi(t)$ is the cyclical
angle between the LOS $\hat n$ and the Earth-SSBC vector. 

The  last effect results from distortion of the scattered image
of the source caused by wavefront curvature,  which 
modulates the integrated flux by a factor $ G = G_x G_y$, where $G_{x,y}$ 
are ``gain'' variations 
related to the second phase
derivatives of the refractive phase in the $x,y$ directions.  
For simplicity,  we assume that
the scattered image would be circularly symmetric in the absence of
wavefront curvature ($G_{x,y}=1$), 
yielding a mean delay $\taudbar^{\rm (iso)}$. 
For $G_{x,y}\ne 1$ the mean delay becomes
\be
\tPBFRISS = \frac{1}{2} \taudbar^{\rm (iso)} \left(G_x + G_y\right),
\ee
as shown in Appendix~\ref{app:screens}.    
Since refractive flux variations are
typically tens of percent but can extend to 100\% or more,  this timing
modulation can be a large fraction of the mean PBF delay. 
The variations induced by the gains
scale as  $G_{x,y}^{-1}-1 \propto \lambda^2$.   Combined with
the $\lambda^{4.4}$ scaling of $\taudbar$, the refraction-induced
variations in $\taud$ are very strongly frequency dependent. 
Variable pulse broadening on a time scale
of $\sim 60$~to~$200$~days was reported for the MSP B1937+21
at the 20\% level \citep{c+90, r+06} that has been attributed to
refractive modulation of diffraction, as described here.    

\section{A Signal Model For Pulse Arrival Times}

We combine the 
physical TOA errors highlighted in \S\S~\ref{sec:achromatic}-\ref{sec:chromatic}
into groups that are convenient for discussing the overall timing error budget
and for fitting to multi-frequency arrival times.
The total TOA perturbation has two terms,
\be
\dtTOA = \dtWHITE + \dtSLOW,
\ee
one with white-noise statistics and
a second that varies  slowly enough that it is
typically correlated between observing epochs.  
A careful analysis of the errors must also
consider the frequency dependences of the various terms.  
The white-noise term is
\be
\dtWHITE = \dtJ + \dtPBFDISS + \dtRN.
\label{eq:white}
\ee
The slow term is made up of strongly chromatic contributions,
\be
\dtSLOW = 
	  \dtDM 
	+ \dtPBF 
	+ \dtPBFRISS
	+ \dtAOA
	+ \dtAOABary
	+ \dtDMnu
	+ \dtRM. 
\label{eq:slow}
\ee

The two terms (white and slow) are treated very differently in
any arrival time analysis.   The white-noise terms 
are unavoidable because they cannot
be fitted for and subtracted from arrival times.  
The slow terms, unlike the white terms,
 can potentially be removed through 
appropriate multi-frequency fitting because there is a finite number
of contributing terms that have different frequency scalings.  
Timing noise and gravitational perturbations also produce slow
variations but are achromatic.

We now discuss the white-noise and slowly varying  terms in detail.

\subsection{White Noise Contributions to Timing Errors}

Pulse-phase jitter, the finite-scintle effect and radiometer noise
are typically uncorrelated between data sets used to estimate
individual TOAs.  Jitter is weakly chromatic, while the other
two contributions are strongly chromatic.  For radiometer
noise, this is due partly to the contribution of Galactic synchrotron
noise at frequencies less than about 0.5~GHz and to the
typical steepness of pulsars' radio spectra.   
However, when interstellar scattering broadens the
pulse at low frequencies, the radiometer-noise error is enhanced  
according to Eq.~\ref{eq:RN} because the pulse-width increases and
$\SNR_1$ decreases.
Scintillations are strongly chromatic.

\begin{figure}[h!]
\begin{center}
   \includegraphics[scale=0.40, angle=0]
	{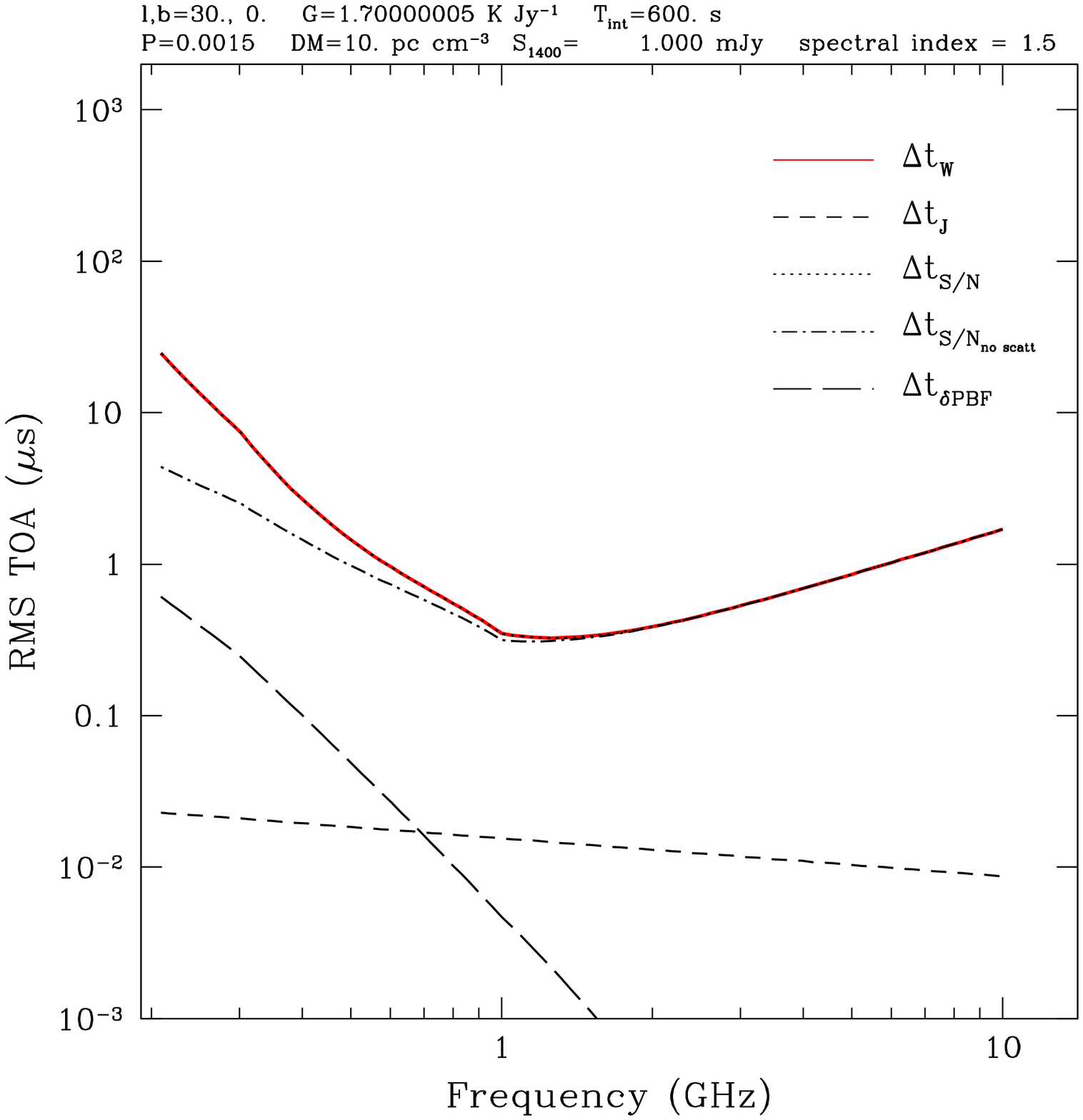}
   \includegraphics[scale=0.40, angle=0]
	{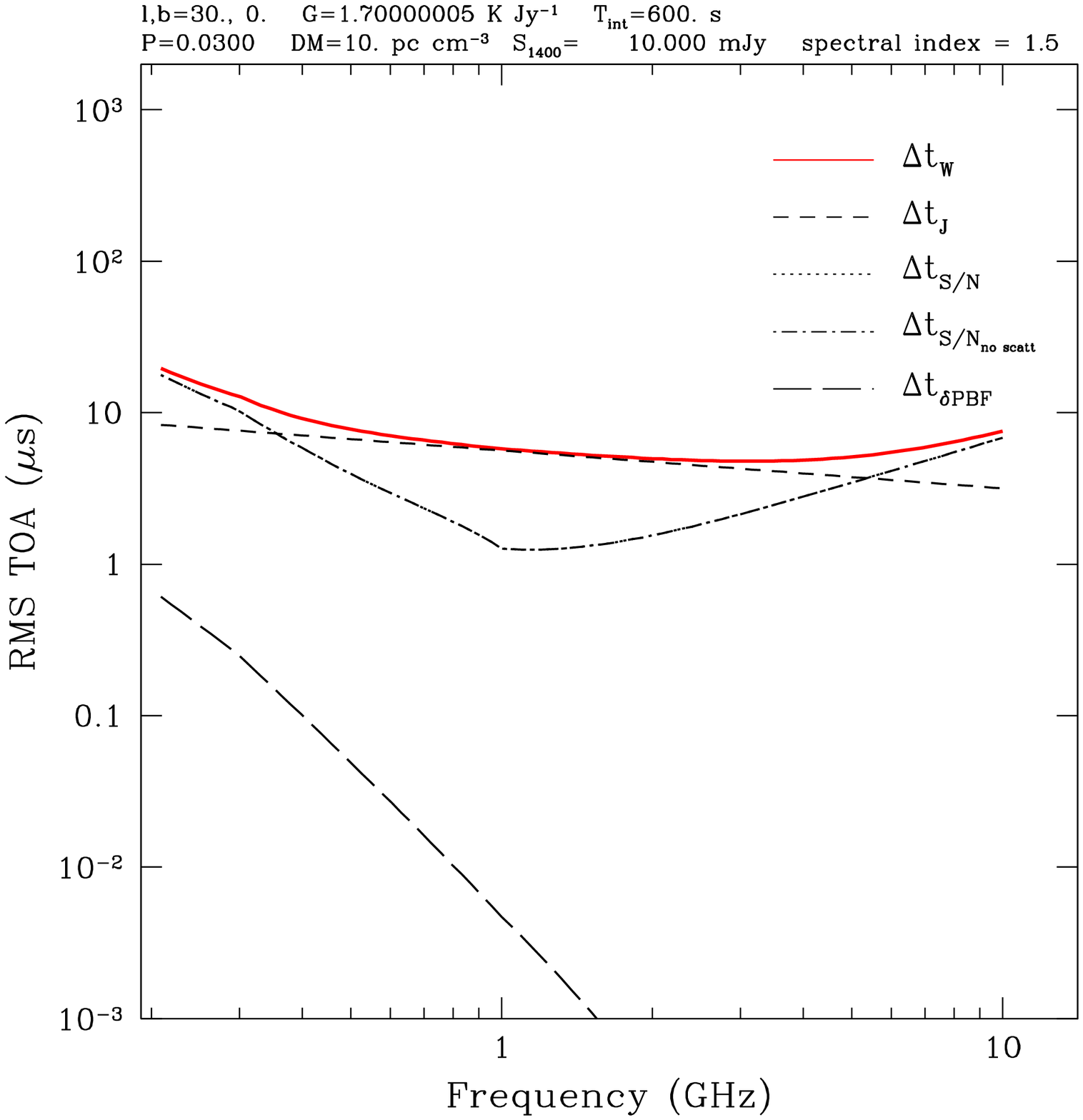}
   \includegraphics[scale=0.40, angle=0]
	{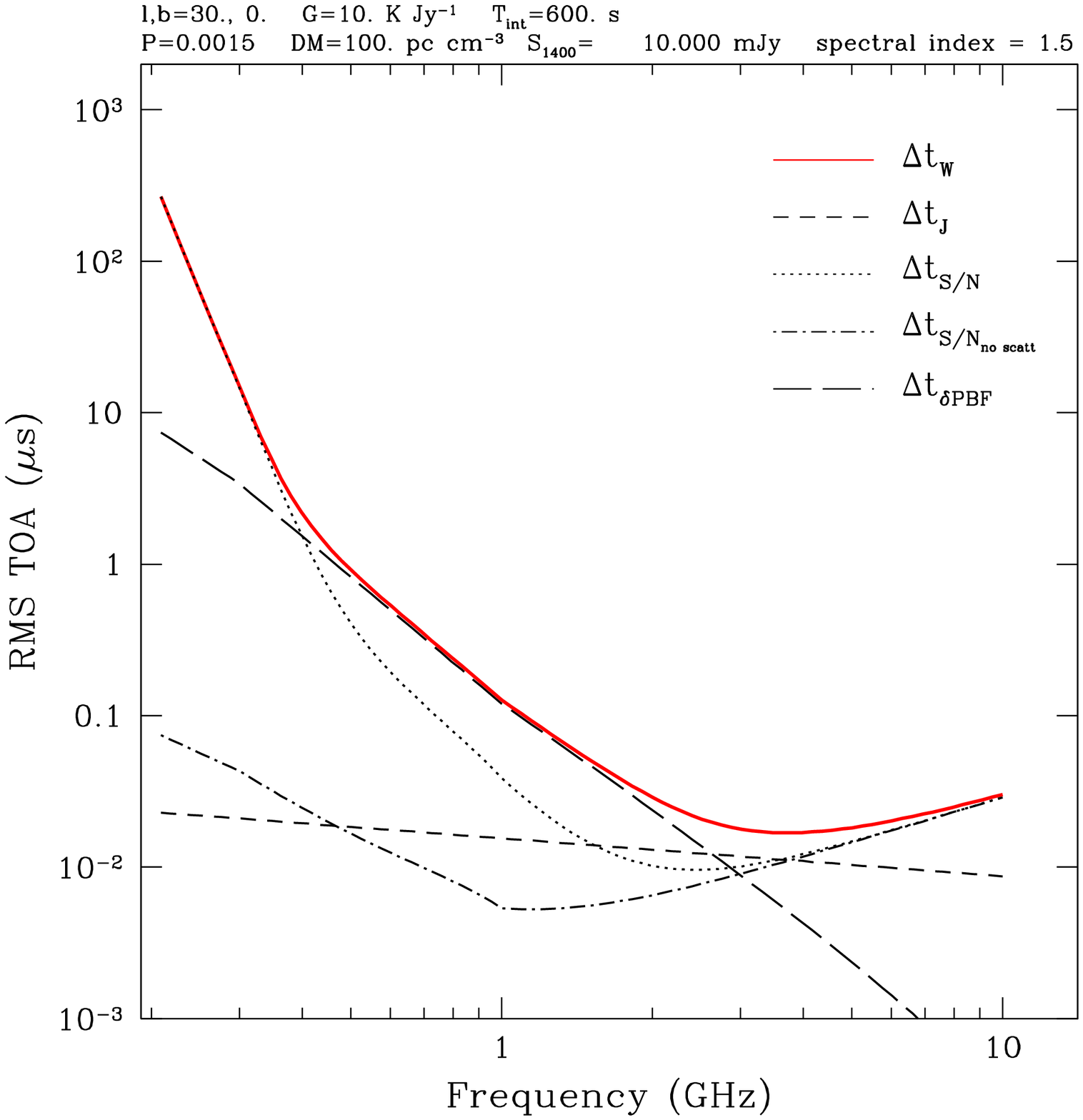}
   \includegraphics[scale=0.40, angle=0]
	{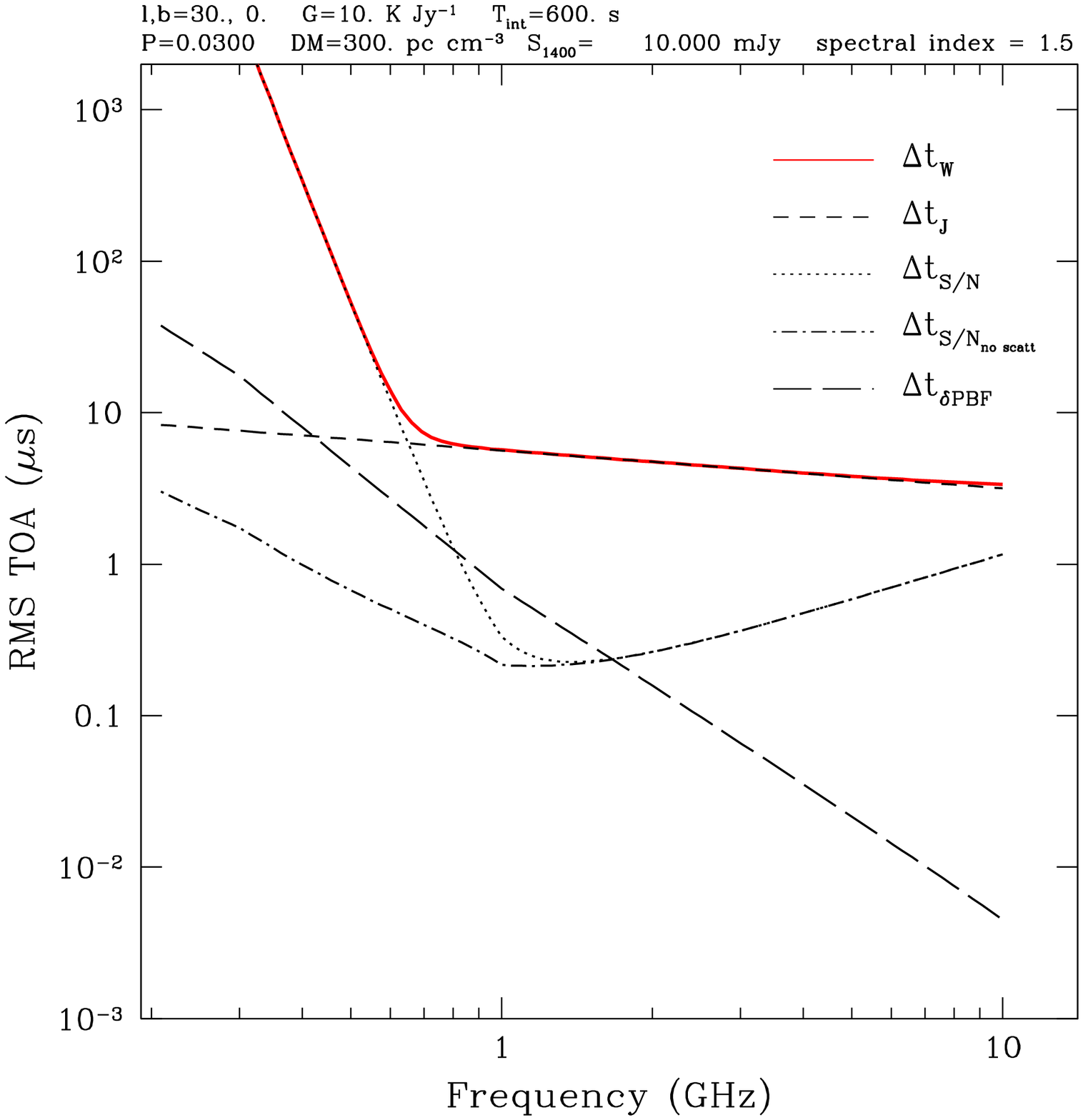}
   \caption{
   Rms timing errors that contribute to the white-noise
   term in Eq.~\ref{eq:white}  are shown plotted against frequency.
   The top two curves are calculated for a 100-m  class (GBT-like) telescope
   with gain $G = 1.7$~K~Jy$^{-1}$ while the bottom curves are
   for an Arecibo-like telescope
   with large gain, $G = 10$~K~Jy$^{-1}$. Flux densities quoted below are
   for 1.4~GHz and scale with frequency as $\nu^{-1.5}$.  The bandwidth
   ratio $B/\nu$ is 0.05 for frequencies below 0.3~GHz, increases linearly
   to 0.2 from 0.3 to 1~GHz, and stays at 0.2 for higher frequencies.
   Solid (red) curves: the total white-noise contribution to the TOA error.
   Dot-dash curves: $\dtRN$ calculated using the intrinsic pulse width;
  	these are underestimates of the true TOA error from radiometer noise.  
   	Radiometer noise includes a constant temperature added to 
   	the 408~MHz background temperature from \citep{h+82}. 
        The intrinsic pulse shape is Gaussian in form with  a 
        duty cycle of 0.03 at 1.4 GHz that scales as $\nu^{-0.2}$.
        The duty cycle also scales with spin period as $P^{-0.4}$
        for periods longer than 2~ms but are allowed to be small (0.02)
	for shorter periods to model those MSPs with
	very narrow pulses.
        The pulse width used to calculate the contribution from
        radiometer noise ($\dtRN$) includes the effects of interstellar
        pulse broadening calculated using the NE2001 electron-density model. 
   Dotted curves: the radiometer-noise timing error ($\dtRN$) calculated 
   	using the apparent pulse width that includes interstellar scattering 
   	broadening;
   Short-dashed curves:  The contribution from pulse-phase-jitter ($\dtJ$)
   	using a jitter parameter $f_J = 1/3$.
   Long-dashed curves: Timing errors from the finite number of
	scintles ($\dtPBFDISS$).
   	The scintillation time used to calculate $\dtPBFDISS$ 
	 is based on a fiducial  transverse velocity of 100~km~s$^{-1}$.
   (Top left:) 
	Millisecond pulsar ($P=1.5$~ms) that is nearby 
	with DM=10~$\DMu$;   
	the TOA is noise limited because
	the flux density (1~mJy) and the telescope gain (1.7~K~Jy$^{-1}$)
	are low. 
   (Top right:) 30-ms pulsar that is nearby but bright enough that
	its timing is jitter dominated for most of the frequency range.  
   (Bottom left:) Millisecond pulsar  with moderate DM = 100~$\DMu$  
	and high enough flux density (10~mJy) observed
	with high gain (10~K~Jy$^{-1}$) so that the timing is
	dominated by the DISS finite-scintle effect.
   (Bottom right:) 30-ms pulsar  with large DM (300~$\DMu$	
	and large flux density (10~mJy) observed with high gain such
	that the timing errors are jitter dominated at high frequencies
	but are noise limited at low frequencies owing to 
	pulse broadening from interstellar scattering.
   \label{fig:rmswhite}
   }
\end{center}
\end{figure}

In Figure~\ref{fig:rmswhite} we show the three white-noise terms for
four different sets of pulse period and \DM.  
The salient features of the cases shown in the figure are that 
\begin{enumerate}
\item For fixed observing time $T$, the phase-jitter effect is small for 
MSPs that have narrow pulse widths and for which a TOA is calculated
from a large number of pulses.
The opposite is true for strong, long-period pulsars, which will have jitter
dominated TOA errors at frequencies $\gtrsim 0.4$~GHz.  
\item The radiometer noise term $\dtRN$ dominates 
other TOA perturbations for weak pulsars (or observations
with small telescopes).  At low frequencies, the timing error
is enhanced by pulse broadening from ISS, which also reduces the
peak flux density. 
\item For strong pulsars, the TOA error is dominated by 
radiometer noise at low frequencies because of the large Galactic
background  and by pulse jitter ($\dtJ$)
at high frequencies.  In some cases (e.g. the bottom-left panel
in Figure~\ref{fig:rmswhite}) at intermediate frequencies
$\sim 0.5 - 2$~GHz, the finite scintle effect ($\dtPBF$) plays a role.
\item  At higher frequencies,
the TOA errors of weak, short period pulsars are dominated
by radiometer noise unless high-gain telescopes (Arecibo, SKA Phase 1, SKA) 
are used.
\end{enumerate}

The primary recourse
is to reduce the amplitudes of the three contributing
terms by increasing the averaging time and  the bandwidth and
by improving the sensitivity of the telescope with a larger
collecting area or reduced system noise temperature.   
As a function of frequency $\dtRN$ and 
$\dtPBFDISS$  are typically uncorrelated
(in the latter case for bandpasses that are separated by 
at least one scintillation bandwidth).
The jitter term $\dtJ$ is correlated even between widely spaced
frequencies because
single pulse structure is typically seen over octave bandwidths, though
not without some evolution in frequency.   A 
multifrequency analysis may be useful for correcting
or reducing pulse phase jitter owing to this partial correlation.

\subsection{Slow, Chromatic Variations}

The slow terms in Eq.~\ref{eq:slow} are all caused by plasma effects
and are therefore strongly chromatic.
From left to right, the terms 
scale with frequency  approximately  as
$\nu^{-2}, \nu^{-4.4}, \nu^{-X}, \nu^{-4}, \nu^{-2},
\nu^{-23/6}$ and $\nu^{-3}$.
We emphasize that such scalings are idealized and likely differ
between lines of sight and also for particular realizations
of sampled data.

In the absence of multipath scattering, 
the DM term scales as $\nu^{-2}$ with departures 
that require very high electron densities or frequencies near
the plasma frequency $\propto \nu^{-4}$ \citep{tzd68, c70}. 
However, multipath scattering introduces an averaging scale on the
DM perturbation that adds the  small term with $\nu^{-23/6}$ frequency
dependence, as discussed in \S~\ref{sec:screenave}.

The $\nu^{-4.4}$ scaling of $\dtPBF$ assumes a Kolmogorov wavenumber
spectrum in the moderate scattering regime.  In super-strong scattering
the exponent becomes $-4$ because the diffraction scale is much less
than the inner scale.   This regime will apply for high-DM pulsars
observed at low frequencies. 
Other departures from the $\nu^{-4.4}$ scaling
for pulse broadening mentioned earlier \citep{l+01, lo+04}
indicate that the ISM has more complex structure than
in idealized models.

The scaling of the AOA term $\propto \nu^{-2}$ also requires discussion.
For discrete structures in the ISM that refract the diffracted radiation
from a pulsar,  this scaling is appropriate.  
However, the scaling differs for density
irregularities from a Kolmogorov spectrum 
because the range of length scales contributing to the refraction
also varies with frequency.   For a Kolmogorov spectrum, refraction effects
are small, however, so it is appropriate to model the AOA term
as we have, given that there is evidence for the presence of
non-Kolmogorov refracting structures in the ISM
\cite[][]{h+05}. 

The frequency scaling of the RISS modulation of pulse
broadening is not a simple power law because the dependence
on the gains $G_{x,y}$ and on any anisotropy in the diffraction
is not simple.   
However, in the following we assume an approximate scaling law, 
$\nu^{-X_{\delta \rm PBF}}$
for this term that applies over a restricted frequency range.

The RM term is likely too small to include in any actual fitting model
for most lines of sight.  It is conceivable that a pulsar of interest
will be discovered for which profile splitting is important because
the RM is large.   The actual scaling law of the timing perturbation
may differ from the generic $\nu^{-3}$ scaling because any effect will
depend on how TOAs are calculated from the measured pulse profiles
in the various Stokes parameters.   

We write terms using values
at a fiducial frequency, $\nu_0$. 
In approximate order of decreasing amplitude we have,
using $z\equiv \nu/\nu_0$,
\be
\dtSLOW(t, \nu) &=& 
	{\dtDM}_0  z^{-2}
	+ {\dtAOABary}_0 z^{-X_{\rm AOA}}
	+ {\dtPBF}_0 z^{-X_{\rm PBF}}
	+ {\dtAOA}_0 z^{-2X_{\rm AOA}}
	\nonumber \\
	&& 
	+ {\dtPBFRISS}_0 z^{-X_{\delta\rm PBF}}
	+ {\dtDMnu}_0 z^{-X_{\rm DM,\nu}}
	+ {\dtRM}_0 z^{-3},
\label{eq:slow2}
\ee
where, for example, for a Kolmogorov medium in strong scattering, 
$X_{\rm PBF} = 4.4$, 
$X_{\rm AOA} = 2$, 
$X_{\rm DM,\nu} = 3.8$
and 
$X_{\delta\rm PBF}$ has a number of possible values as described earlier.

\section{Mitigation of Timing Errors from Non-dispersive Chromatic Effects}
\label{sec:mitigate}

We have identified salient timing errors that are introduced by 
intervening plasmas. The primary question
is how to contend with them when precision timing is
required, as for the detection of gravitational waves with pulsars. 
We discuss two broad approaches that are based on fitting ``raw''
arrival times vs. frequency $\nu$ (as opposed to first correcting 
TOAs for one or more effects with subsequent fitting).

An augmented version of Eq.~\ref{eq:tnu1} for the TOA includes
a non-dispersive, chromatic term, $t_C$,
\be
t_{\nu} = t_{\infty} + t_{\DM}(\nu) + t_C(\nu) + \tWHITE(\nu). 
\label{eq:tsimple}
\ee 
To simplify the notation, we have dropped 
the `$\Delta$' used previously to represent fluctuations.
In the following we consider multifrequency TOAs at a given 
epoch.   Consequently we include in $\tWHITE$ only the radiometer noise
and scintillation effect.   The jitter term in this context is 
considered to be frequency independent (see \S~\ref{sec:jittermain}).
When using the variance of the white noise in
weighted least-squares fitting across frequency, we exclude 
the jitter term for this reason. However, the jitter term does contribute
to epoch-to-epoch timing errors, as previously discussed.

\subsection{Ignore Refraction and Scattering Effects}

One approach is to simply ignore all effects other than the dispersion delay.
Most arrival-time analyses to date include the dispersive
term $\dtDM$ as the only chromatic perturbation. Such programs 
use dual or multiple-frequency TOAs to estimate DM at each epoch
or use ancillary measurements of DM to
correct TOAs to infinite frequency at the SSBC.   
This procedure allows scattering 
to contaminate the estimated TOA \citep{fc90}. 
To be successful, removal of dispersion delays only  
requires additional errors to be smaller than some target precision. 
As in Figure~\ref{fig:regimes} for the case of pulse broadening,
restrictions can then be made on the combination of
 observation frequencies and DMs of a pulsar sample that can
satisfy the error budget.  

Following the same approach discussed in \S~\ref{sec:dDM} and
in Appendix~\ref{app:lsq}, we calculate the error on
the corrected TOA, $t_{\infty}$, when the $t_C$ term is present
in the data but only the $t_{\DM}$   term is  included in the fit.  
For example we consider the simple case where that $t_C$ includes only 
one term instead of the many terms of Eq. ~\ref{eq:slow2} and that
it scales as $t_C(\nu) = a_C\nu^{-X}$, 
(where we drop any subscript on the exponent, $X$). 
The two parameter fit for $t_{\infty}$ and $\DM$
yields a systematic  error for $t_{\infty}$ (in addition to 
the random, standard error) 
that can be written as
\be
\delta t_{\infty} = 
		- a_C R_{t_{\infty}},
\ee
where $R_{t_{\infty}}$ is a coefficient defined in 
Eq.~\ref{eq:dtinfty1}.  
For $X=4$, a 2:1 range of frequencies,
and frequency channels ranging in number from 2 to 1024,
we find a variation 
$R_{t_{\infty}} \approx 0.3\pm0.05$.
Thus a fraction $1 - R_{t_{\infty}}\sim 70$\% of the 
non-dispersive delay $a_C$ is absorbed
in the fit for \DM, yielding an estimate for $t_{\infty}$ that is
too early by $\sim (0.3\pm0.05)a_C$.  
For a 4:1 frequency range, $R_{t_{\infty}} \approx -0.1$, so
90\% of $a_C$ is absorbed.   However, for a 1.5:1 range,
only about 50\% is absorbed in the fit. 
Referring to Figure~\ref{fig:regimes} 
for a 2:1 frequency range, 
the  maximum timing error $\sigma_t$ from this effect alone requires that
$a_C < \sigma_t / R_{t_{\infty}}$ or, since 
$a_C \equiv t_C(\nu_1)\nu_1^X$,
\be
t_C(\nu_1) <  \nu_1^{-X}R_{t_{\infty}}^{-1} \sigma_t.
\ee
For $\nu_1 = 2$~GHz, $\sigma_t = 0.1~\mu s$ and using
$R_{t_{\infty}} = 0.3$, we need
$t_C(2~{\rm GHz}) < 19~ns$. 

\begin{figure}[h!]
\begin{center}
   \includegraphics[scale=0.40,angle=0]{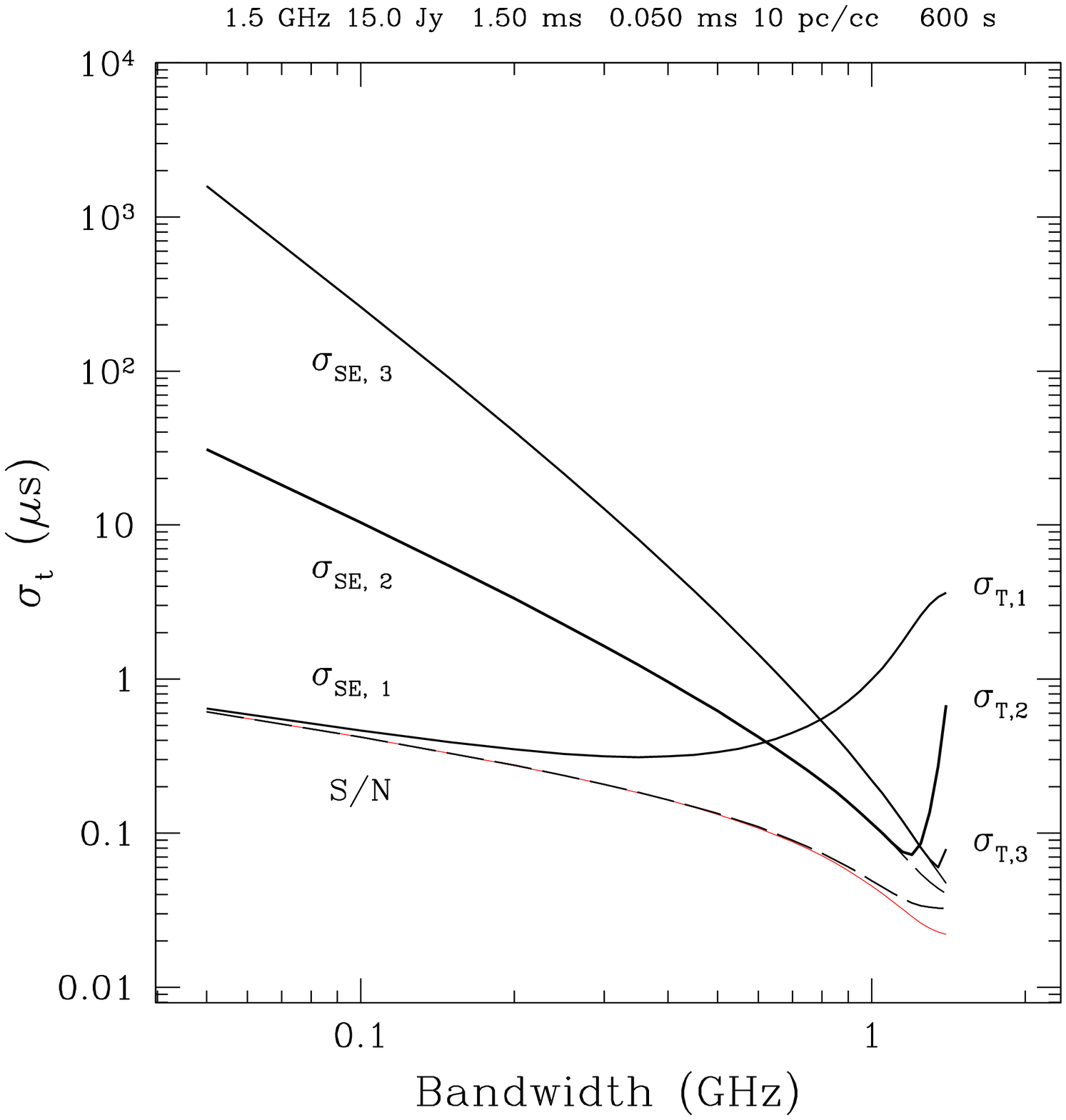}
   \includegraphics[scale=0.40,angle=0]{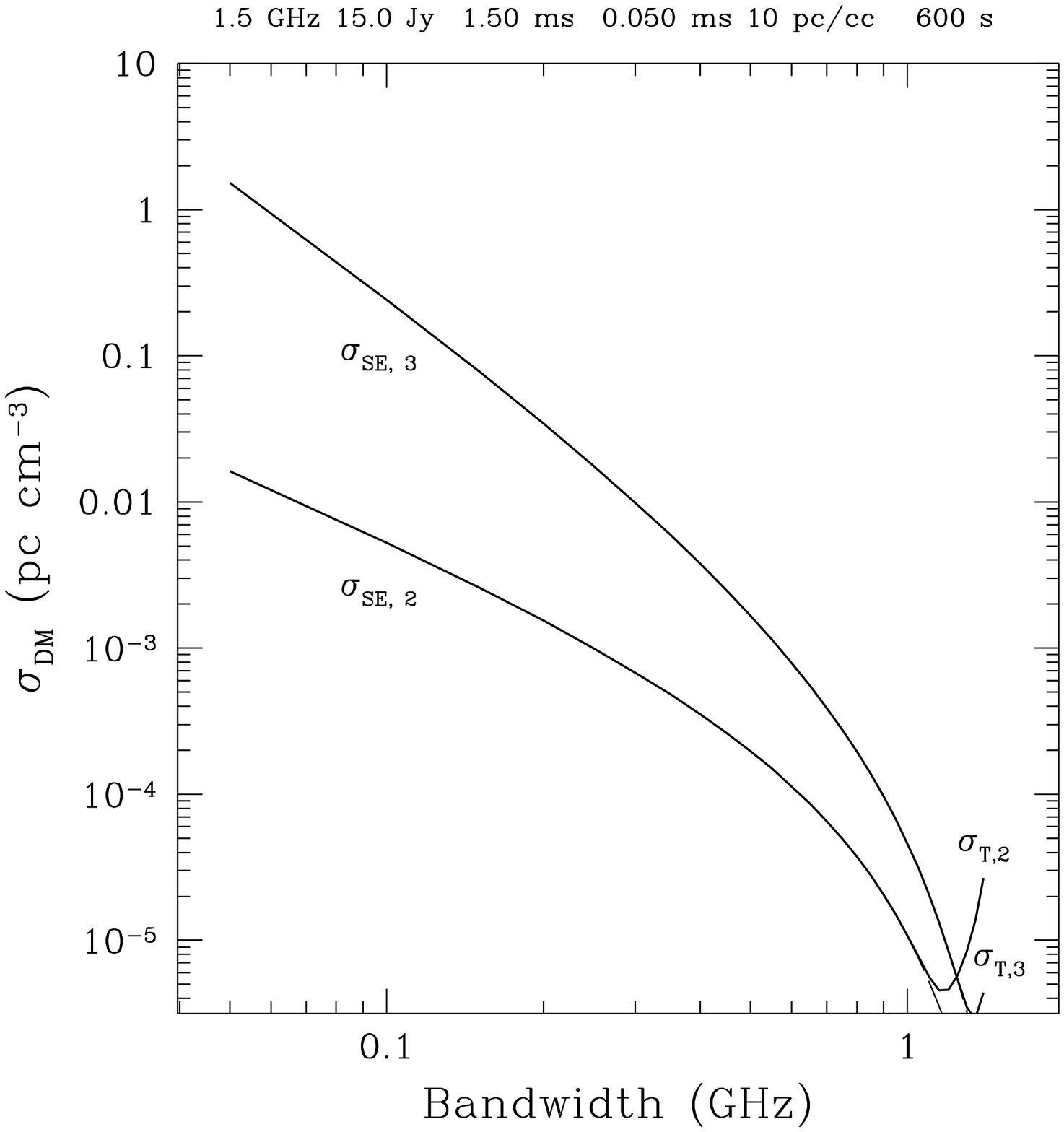}
   \caption{
	Left:
	Standard and systematic errors after 
	fitting multifrequency TOAs across the designated bandwidth  
	with a maximum frequency of 1.5~GHz.
	These errors apply to cases where the only chromatic terms included are
	the dispersive term $t_{\DM}$ and an additional term $t_C\propto \nu^{-X}$
	with $X=4.4$ 
	that approximates pulse broadening or one of the refraction terms 
	discussed in the text. 
	The curves apply to a 600~s observation using a 100-m class telescope 
	of a pulsar with $P=1.5$~ms and $\DM\ = 10~ \DMu$ in the direction $\ell = 45^{\circ}$,
	$b = 0^{\circ}$.  The pulse width is $50~\mu s$. 
	Dashed lines labeled $\sigma_{\rm SE, k}, \, k=1,2,3$ are the standard errors for
	one, two and three-parameter fits as discussed in the text.  
	%
	Solid lines labeled $\sigma_{t,k}$ are total errors that quadratically sum standard and
	systematic errors.   For $k=1$, the systematic error includes an error in dispersion measure,
	$\delta\DM = 10^{-3.5}~\DMu$ along with 
	pulse broadening calculated using NE2001.
	For $k=2$, the systematic error is solely from pulse broadening because $\DM$ is fitted for.
	For $k=3$, an error $\delta X = 0.44$ (10 \%) is assumed in the exponent of the 
	pulse-broadening scaling law.
	%
	The curve labeled ``S/N'' is the standard error
	resulting solely from radiometer noise.
	It scales as $B^{-1/2}$ for small bandwidths but steepens
	as larger bandwidths include frequencies where the
	flux density is higher. 
	The difference between
	the total error and the standard error for the 
	two-parameter fit is caused by the scattering term, $t_C$.   
	This example shows that a 3-parameter fit can remove
	much of the scattering when large bandwidths 
	are used, allowing pulsars with large DMs to be included in 
	a pulsar timing array
	 Right: Similar plots for errors in DM when it is included in two and 
	three-parameter fits.  
	\label{fig:sigB1}
   }
\end{center}
\end{figure}

A more detailed analysis is shown in 
Figure~\ref{fig:sigB1} with calculated results for
an MSP with $P=1.5$~ms with low DM (10~$\DMu$)  
with Galactic coordinates $\ell, b = 45^{\circ}, 0^{\circ}$
and a pulse width of $50~\mu s$.
The period-averaged flux density is
1~mJy at 1.4 GHz and scales with frequency as $\nu^{-2}$.  
The NE2001 model is used to estimate
the distance and  pulse broadening from scattering. 
In the left hand panel,
the rms timing error is plotted against bandwidth $B$
for the case of an MSP observed with a 100-m class
telescope.  The right-hand panel shows the corresponding
errors for \DM. 
The example shown in the figure  applies to observations using 
100-m class telescopes.
The highest frequency in the example is 2~GHz  so 
the measurements and fit therefore  cover the frequency range $[2-B, 2]$~GHz 
for each bandwidth plotted. The left-hand panel shows standard errors for
fitting functions with one to three parameters 
($t_{\infty}, \DM$ and $a_C$) and systematic errors for the 
three cases.  Other details are described in the caption.
The two-parameter fit is relevant to this section where
pulse broadening is ignored. 
Salient features of the figure include:
\begin{enumerate}
\item For the $k=2$ fit for $t_{\infty}$ and $\DM$ the total error minimizes
	at $B\approx 1.4$~GHz as pulse broadening becomes important at
	the lower frequencies in the band. 
\item When pulse broadening is ignored in any fitting, the best
	approach is to use a two-parameter fit if \DM\ is not known
	to sufficent precision to warrant a single-parameter fit. 
\item For small bandwidths, standard errors 
	(dashed lines labeled $\sigma_{\rm SE, k}, \, k=1,2,3$)
	become increasingly larger in going from one to three-parameter fits
	because the weak dispersive and scattering terms are not well modeled.
        The timing error scales as $B^{-1/2}, B^{-3/2}$ and $B^{-5/2}$,  respectively, when the TOA error is dominated
        by radiometer noise, as it is for small bandwidths. 
	For $B\ll\nu_1$ the ratio
	$\sigma_{\rm SE, 2} / \sigma_{\rm SE, 1} \approx \sqrt{3}\nu_1/B$
	and is a factor of 69 for the smallest $B$ plotted in 
	Figure~\ref{fig:sigB1}.
	The penalties for fitting for \DM\ and $t_C$ can be avoided if 
	the dispersion
	measure and pulse broadening are known to sufficient accuracy from 
	ancillary measurements, but both of
	these quantities vary with epoch and must be monitored.   
\item Foreshadowing results in the next section,
	the three parameter fit improves the TOA error but only
	if the scaling exponent $X$ is known.   The upturn in
	$\sigma_{T,3}$ is caused by a
	10\% uncertainty in $X$, $\delta X = 0.44$, illustrating
	the requirements on prior knowledge of $X$.   
	Nonetheless the improvement is by a factor  of about 10 
	and can be greater if $X$ is known to better precision.  
\item The best achievable timing precision corresponds to 
	$\sigma_{T,3}$ and is about 0.1$~\mu s$ for the 
	600-s observation shown and with a 1.8~GHz bandwidth.
\item Brighter pulsars will shift the curves downward over most, but not
        all of their plotted range. 
\end{enumerate}
We emphasize that the results shown in Figure~\ref{fig:sigB1}
and in the next two figures
do not include all refraction and diffraction effects, so
they must be taken as illustrative and optimistic for
considerations of ISM effects alone. 

\subsection{Aggressive Removal of  Refraction and Scattering Effects}

If non-dispersive effects are tackled  more aggressively,
a larger number of pulsars can be included in a timing program because
a wider range of frequency-DM combinations can satisfy a given 
error budget.  


Consider the case where,
along with $t_{\infty}$ and $t_{\DM}$, 
we include $t_C$ in the fitting function as a power-law
$t_C = a_C \nu^{-X}$. 


This approach does not require knowledge of $C_1$ or measurement
of the scintillation bandwidth $\dnud$ as discussed in
\S~\ref{sec:corr} but it does
require assumption of a value for the exponent, $X$. 
For now, we consider $X$ to be known and we consider $t_C$ to describe
only pulse broadening.





\begin{figure}[h!]
\begin{center}
   \includegraphics[scale=0.40, angle=0]{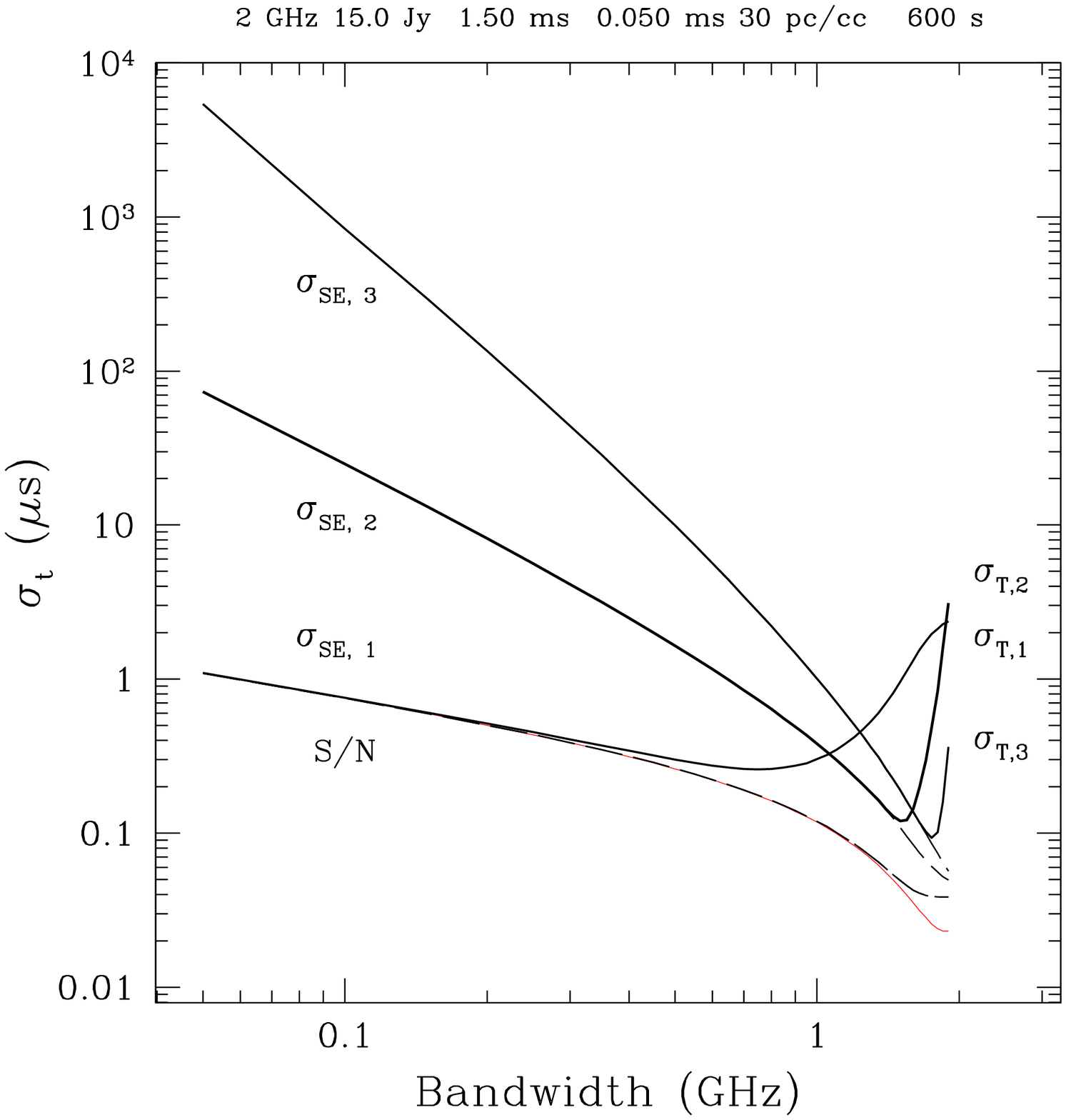}
   \includegraphics[scale=0.40, angle=0]{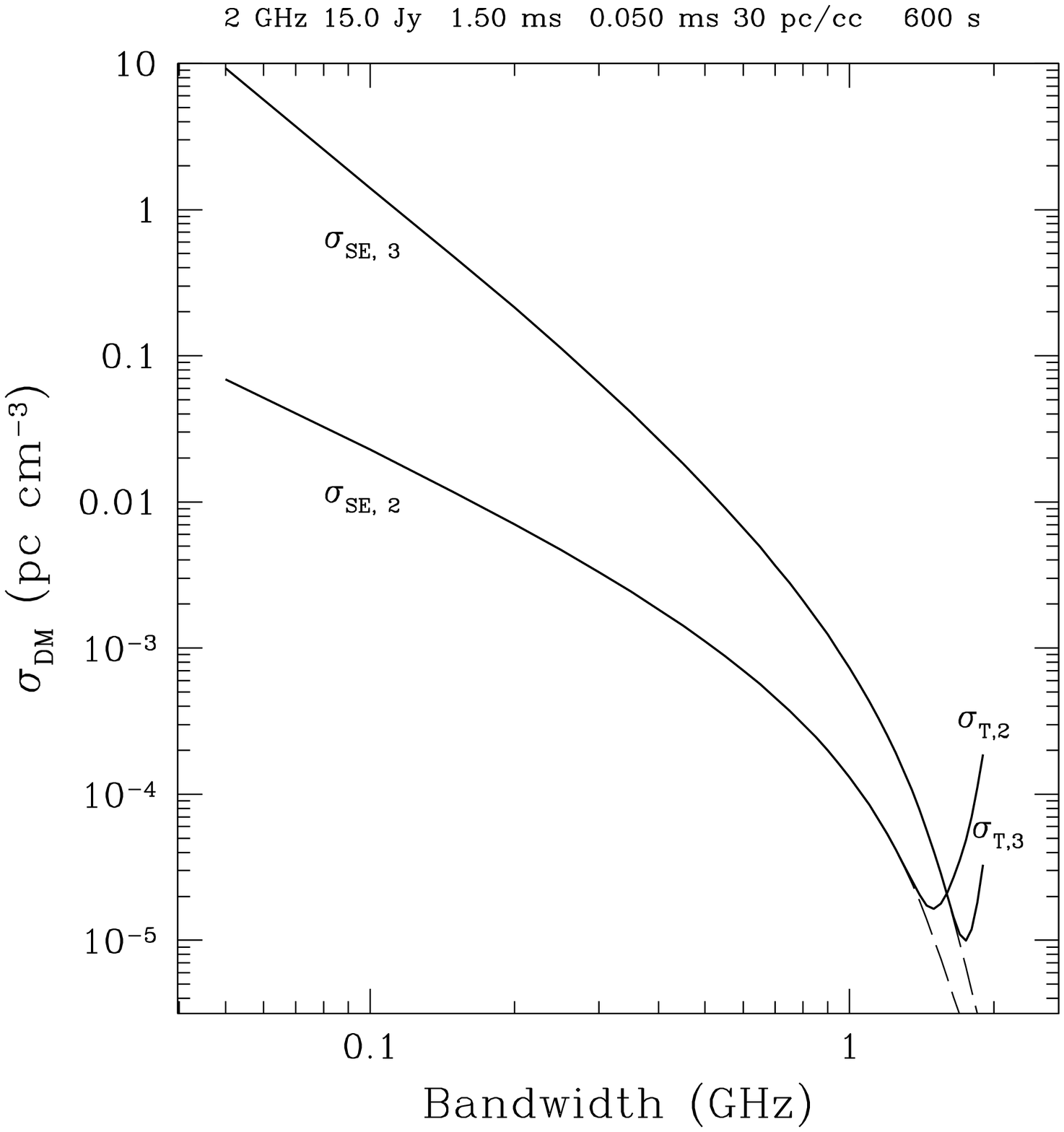}
   \caption{
	Plot of standard and systematic errors for multifrequency fits to 
	arrival times in the same format and for the same pulsar parameters  
	as in Figure~\ref{fig:sigB1} except 
	for a dispersion measure $\DM = 100~\DMu$  and a maximum
	frequency of 2~GHz. 
	The left-hand panel is for a 
	100-m class telescope with $\Ssys = 15$~Jy  and the right-hand panel
	is for an Arecibo-class telescope with $\Ssys = 4$~Jy.
	\label{fig:sigB2}
	}
\end{center}
\end{figure}

Figure~\ref{fig:sigB2} shows results similar to those
in Figure~\ref{fig:sigB1} for observations with
$\nu_1 = 2$~GHz of
a 1.5~ms pulsar at $\DM = 100~\DMu$ using a 100-m class
telescope (left panel) and a 300-m class telescope.
The latter case corresponds approximately to the Arecibo telescope,
the Chinese FAST telescope, and the proposed Phase 1
of the SKA. 

The standard errors shown in the figure are larger for
the 100-m class telescope in accord with its smaller
sensitivity by about a factor of four.   The total
error from the two-parameter fit minimizes for 
$B\approx 0.5$ GHz and $B\approx 0.4$ GHz   
for the smaller and larger telescopes, respectively,
the upturn being caused if there is no correction
for pulse broadening.  A similar upturn occurs
for the three parameter fit if there is an error
in the scaling-law exponent $\delta X = 0.44$ (10\%).
If there is no error, the best rms error is for
a bandwidth $B=1.6$~GHz and is approximately 2.5
times better for the larger telescope. 

The conclusions that can be made from this particular case
are that aggressive treatment of pulse broadening can
drastically improve timing precision of moderate-DM
pulsars and thus allow more pulsars to be included
in a pulsar timing array than otherwise would be the
case.  Without addressing pulse broadening, there is a 
maximum usable bandwidth that is less than 1~GHz 
and the timing precision $\sim 1~\mu s$ is not
competitive for PTAs unless more observations
are made during a year. 

\begin{figure}[h!]
\begin{center}
   \includegraphics[scale=0.40, angle=0]{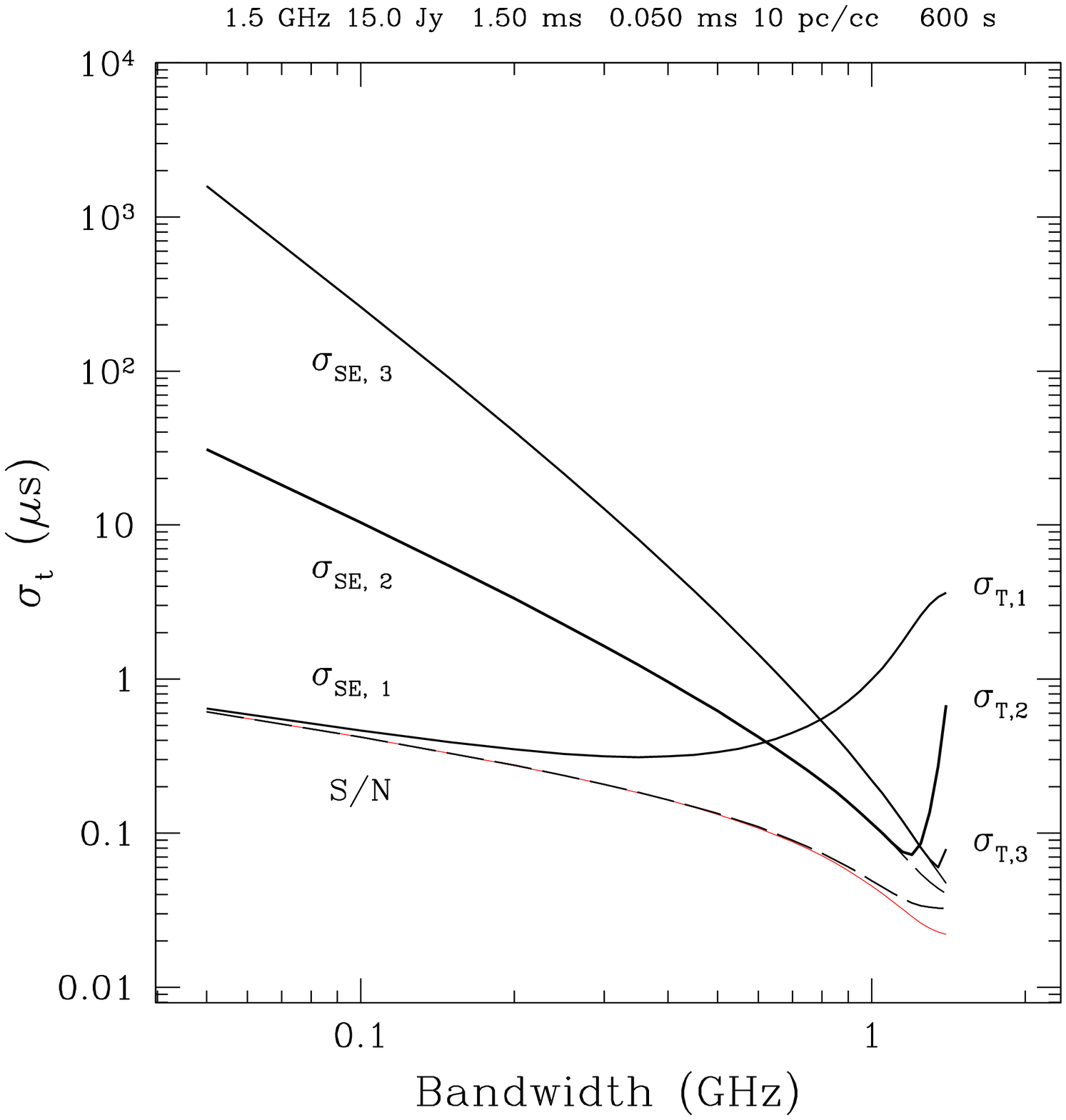}
   \includegraphics[scale=0.40, angle=0]{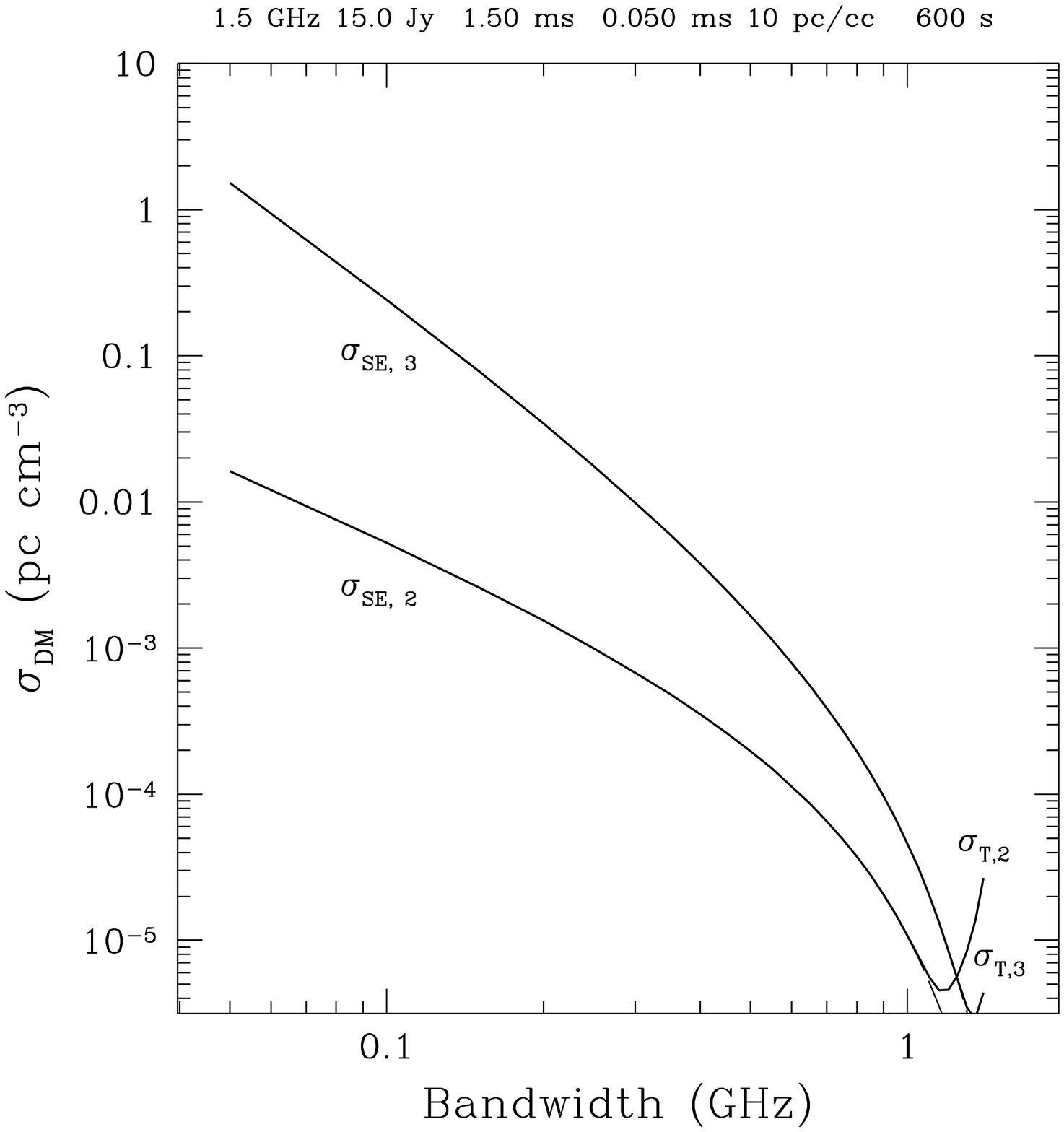}
   \caption{
	Plot of standard and systematic errors for multifrequency fits 
	to arrival times in the same format and for the same pulsar 
	parameters  as in Figure~\ref{fig:sigB1} except for 
	a dispersion measure $\DM = 500~\DMu$  and an upper frequency
	of 5~GHz.  
	The left-hand panel is for a 
	an Arecibo-class telescope with $\Ssys = 4$~Jy) and 
	the right-hand panel is for the full SKA with $\Ssys = 0.3$~Jy.
	\label{fig:sigB3}
	}
\end{center}
\end{figure}

Figure~\ref{fig:sigB3} compares observations
of a high \DM\ pulsar  ($500~\DMu$)
at high frequency (5~GHz)  for observations with
an Arecibo-class instrument (left panel)
and an SKA-class instrument (right panel). 
For the SKA we have used a nominal sensitivity
$A_{\rm eff} / T_{\rm sys} = 10^4~{\rm m^2~K^{-1}}$. 
Note that the spin period in this case is 3~ms.
The results indicate that aggressive fitting
for pulse broadening can yield timing precision
that is competitive for PTA applications even
for a high-DM object. 

The examples shown in Figures~\ref{fig:sigB1} - \ref{fig:sigB3} that
include only pulse broadening can be used as a guide for how to
deal with the other chromatic terms in Eq.~\ref{eq:slow2}. 
In total there are eight unknowns to be solved for
at each epoch  when we include the achromatic term
$t_{\infty}$.  However,  the RM term can almost always be
ignored.   The term
$\dtAOABary$ is or almost is  degenerate with the DM term
in its frequency dependence $\propto \nu^{-2}$
and three of the other chromatic terms 
($\dtPBF, \dtAOA$ and $\dtDMnu$) are degenerate with
scalings close to $\nu^{-4}$.    This suggests a minimum
of four unknowns that require four measurements at each epoch over
a frequency range that is adequate to solve for the unknowns.   
Additional terms can be fitted for to potentially improve the
net timing precision. However, examples shown in the figures indicate
that the ``cost'' of extra parameters is high  and that the amplitudes
of additional effects need to be large enough to surpass the cost
in precision.   

\subsection{Comparison with Well-timed MSPs}
\label{sec:welltimed}

We compare our predictions with actual measurements on four 
prominent MSPs to further illustrate timing budget considerations. 
All four objects show TOA precision of 200 to 300~ns in 5~min averages
\citep{hbo06} where, to be explicit, we mean the precision of the
TOA estimate and not necessarily the rms residual over a time span of
years. 
Three objects  (J0437$-$4715, J1713+0747, and J1909$-$3744)
show rms timing residuals that, when integrated over $\sim 1$~hr,
are below 100~ns.   The fourth, 
B1937+21 (J1939+2134), is bright but shows timing noise consistent
with that seen in canonical pulsars, which is caused by torque variations. 
We use flux densities, pulse widths and other parameters from
\cite{hbo06} and other references, as noted, and use parameters
for the Parkes telescope and the  48~MHz bandwidth and  300-s integration
time used  by \cite{hbo06}.  


J0437$-$4715 is the brightest of the four pulsars, with a 
period-averaged flux density $S_{1400} = 137$~Jy.    
Eq.~\ref{eq:RN} combined with the  pulse width $W = 85~\mu s$ and 
the system noise of the Parkes telescope ($\Ssys = 29$~Jy) 
predicts $\dtRN = 13$~ns and $\dtJ = 105$~ns.
This object, not surprisingly, is very heavily jitter dominated.   
The low-DM (2.6~$\DMu$) implies that interstellar effects will be much smaller 
than either of these leading effects. Increases in sensitivity
with larger bandwidths or larger telescopes will not improve the
TOA precision, but longer integration times and a larger numbe of
TOAs will. 

J1713+0747 is less bright and has a larger DM (16~$\DMu$) 
than J0437$-$4715.
For the Parkes telescope, the radiometer noise TOA error
is about 50\% larger than the jitter error (100~ns vs. 65~ns),
but for larger telescopes jitter will dominate.   Diffractive
interstellar scintillations will strongly modulate the signal level    
at most frequencies, making even the Parkes measurements jitter
dominated some of the time.

J1909$-$3744 is one of the best timing pulsars.
The low DM ($10.4~\DMu$) is similar to that shown in 
Figure~\ref{fig:sigB1} and implies that interstellar effects are
small at frequencies above 1~GHz.  
The measured rms residual
is $\sigma_t = 0.3~\mu s$.   
We estimate
$\dtRN \approx 0.41~\mu s$,
consistent with
the curve labeled `S/N' in Figure~\ref{fig:sigB1} (left-hand panel)
for the bandwidth used and after taking into account the small differences
in values for period, width, flux density and system noise as well as
the radio frequency and spectral index.
\cite{b07} notes that the pulsar flux density varies from DISS 
by a factor of 30 and \cite{j+05} exclude TOAs with errors in excess
of $1~\mu s$. The difference between the predicted and actual rms 
residual is attributable to modulations by DISS and by RISS.   
The narrow pulse for this object ($42~\mu s$) implies that pulse
jitter does not make a significant contribution to the observed
measurements ($\dtJ \approx 27~ns$ from Eq.~\ref{eq:jitter1})
unless the scintillated flux becomes very large.  

The fourth object, B1937+21, has the shortest period
($\sim 1.6$~ms)  and narrowest pulse ($30~\mu s$) of the
four objects we consider.  It also has the largest DM ($71~\DMu$)
and is fairly bright.   For the Parkes telescope, the TOAs are
noise dominated with estimates $\dtRN = 56$~ns and
$\dtJ = 19$~ns but will be jitter dominated
for most larger telescopes.    As noted earlier (\S~\ref{sec:jitterknown}),
jitter in this object is ambiguous, displayed in giant pulses
but not in typical pulses, so our estimate of $\dtJ$ is uncertain
and there is the possibility, not yet demonstrated, that timing
precision can be improved by excluding giant pulses from 
timing estimates or by otherwise correcting for them.

\section{Summary and Conclusions}

Table~\ref{tab:synopsis}  summarizes 
how timing precision may be limited  by different causes 
and what the course of mitigation might be.  
We include timing noise intrinsic to the pulsar because many pulsars 
will in fact have timing residuals that are dominated by timing noise
unless it can somehow be removed \cite[e.g.][]{l+10}.

\begin{deluxetable}{llll}
\tabletypesize{\footnotesize}
\tablewidth{420pt}
\tablecaption{\label{tab:synopsis}
Regimes for Pulsar Timing Limitations
}
\tablecolumns{4}
\tablehead{
 \colhead{Limiting Effect}
& \colhead{Spectral Signature}
& \colhead{Regime Specifications}
& \colhead{Mitigation}
\\
& \colhead{in TOAs}
} 
\startdata
Timing noise & red & spin, torque noise & more pulsars in timing array, 
\\
             &     &                    & remove torque variations 
\\
Low S/N & white & weak pulsar, small telescope & larger $T, B$; 
	smaller $S_{\rm sys}$
\\
Phase jitter & white & large S/N & larger $T$
\\
Diffractive scintillations & white & moderate DM & large $T, B, \nu$ 
\\
\quad\quad (Finite scintle effects)
\\
S/N + pulse broadening & white & large scattering & larger $\nu$ or 
		reject pulsar
\\
DM, AOA variations & red & moderate DM objects & multi-frequency fitting
\\
PBF + variations & red & slow changes in $\tau_d$ & multi-frequency fitting
\enddata
\end{deluxetable}


For many objects, timing precision will be limited by a combination
of radiometer noise and pulse-to-pulse phase jitter intrinsic to the pulsar.
We have summarized the levels of jitter in different pulsars as presented
in the literature and we have presented specific results on the
bright MSP, J1713+0747.  We have also shown the relationship
between jitter and pulse-shape stability.    The theoretical best
timing precision is achieved when it is limited by additive white noise
alone and it is straightforward to calculate the rms timing error 
$\dtRN$ using a pulsar's template pulse shape (Appendix~\ref{app:jitter}).  
Comparison of actual TOA errors against $\dtRN$ shows that 
for all pulsars except the MSP B1937+21 \citep{jg04},  the rms phase jitter
appears to be a significant fraction of the pulse width and thus will
limit timing when the signal-to-noise ratio is high.

In the {\it noise dominated regime}, TOA precision may
be improved by  actions that favorably alter any factor in
Eq.~\ref{eq:dtRN}: increasing the sensitivity
(i.e. reducing $S_{\rm sys}$) and the total
bandwidth to increase the  signal-to-noise ratio;
improving dispersion removal to minimize
the effective pulse width $\fwhm$; and increasing the number of pulses
summed. 
In the {\it jitter dominated regime}, however, TOA
precision is improved only by summing more pulses.
Thus in either regime, an increase in telescope time per object is needed
to decrease the timing errors from radiometer noise or jitter. 
As more sensitive instruments are built, more pulsars 
will have jitter-dominated timing, as we have shown in 
\S~\ref{sec:jittermain}.  With the Square Kilometer Array, for example,
most of the known canonical pulsars and the majority of MSPs
will have jitter-dominated TOAs.  Consequently, the SKA can be used
with sub-arrays in order to increase the amount of observing time
per object with the sensitivity of a sub-array set so that the
TOAs are jitter dominated by only a small factor. 

Timing variations from interstellar dispersion and
scattering need to be characterized for each LOS owing to
the richness of structure in the ISM.  Low-DM pulsars
(e.g. $\lesssim 50~\DMu$) are clearly better for precision timing
at frequencies $\sim 1$~GHz but the maximum DM is smaller
for lower frequencies.  For pulsar-timing array applications that use
an ensemble of MSPs,  the best approach is to find
spin-stable MSPs with small dispersion measures.  However, intrinsic
spin noise in MSPs \citep{sc10} may require usage of higher-DM MSPs
and mitigation of interstellar effects.  We have shown that
multiple frequency fitting of TOAs obtained over a large total range
of frequencies can extend the allowed sample of pulsars. 
    Objects in relativistic
binaries are rare enough that future discoveries are likely to include
pulsars with large DMs.   The most extreme case is for pulsars orbiting
the Milky Way's central black hole, Sgr~A*, which will show extreme
levels of scattering with corresponding large timing variations
\citep[e.g.][]{dcl09}.

The dominant interstellar timing error is from variations in DM
followed by diffractive pulse broadening and its variations.
Both of these effects are correctable to vary degree.
Additional interstellar terms imply that timing
measurements at a minimum of four narrowband
frequencies over approximately a 2:1 frequency range.
Better yet --- and now feasible --- are continuous measurements over
 octave-like frequency ranges.    For small dispersion measures. the pulse
broadening time is too small to be determined from TOAs alone even though
it is large enough to influence the arrival times.   
We have outlined a procedure that fits nominal TOAs at multiple frequencies
to remove chromatic effects.  However, other approaches, also mentioned
here, could alternatively remove some of the chromatic effects at
individual frequencies before doing any multiple frequency fits.
Further exploration is needed on using
dynamic spectra and their analysis using either correlation functions or a
secondary spectrum analysis \citep[e.g.][]{hs08} to provide
the necessary timing corrections.   
Other approaches that use phase
information in high resolution sampling 
(e.g. P. Demorest, W. van Straten \& M. Walker, private communication)  
also need to be explored. 

We thank P. Demorest, B. Rickett, D. Stinebring, and M. Walker for useful 
discussions or correspondence and the Nanograv collaboration
and the Lorentz Center
for providing  useful venues for discussions of timing precision.  
JMC had very fruitful and enjoyable conversations 
about timing precision with Don Backer over 
many years that have contributed to this paper. 
This work was supported by the National Science Foundation, which supports the
Arecibo Observatory under a cooperative agreement with Cornell University.


\newpage

\appendix

\section{A. Time-of-Arrival Errors from Pulse Profile Stochasticity} 
\label{app:rn}
\label{app:jitter}

Here we give expressions for the error in time of arrival (TOA) due to
additive radiometer noise and to pulse-phase jitter that is
intrinsic to the pulsar.  

Radiometer noise is additive in the limit of large time-bandwidth product
and causes a minimum error when matched filtering is used to estimate
the TOA.
Let $U(t)$
be the pulse waveform as a function of time and normalized to
unity amplitude.
Also define the autocovariance function of the noise to be
$\rho(\tau) \equiv \langle n(t) n(t+\tau)\rangle/\sigma_{noise}^2$
such that $\rho(0) = 1$  and its characteristic width is 
$\Delta \equiv \int d\tau \, \rho(\tau)$.
For a  signal-to-noise ratio $\SNR$ (pulse maximum/rms noise),
the rms timing error is \citep{dr83}
\be
\dtRN &=& 
\frac{
\left [  \int\int dt \, dt^{\prime}\, \rho(t - t^{\prime})
        U^{\prime}(t) U^{\prime}(t^{\prime}) \right ]^{1/2}}
     {\SNR \int \, dt \, [U^{\prime}(t)]^2 }
=
\frac{W_{\rm eff}}{\SNR_1 \sqrt{N}}
	\left(\frac{\Delta}{ W_{\rm eff}} \right)^{1/2}
= \frac{S_{\rm sys}}
	{S_{\rm peak}\sqrt{2B}} \left(\frac{W_{\rm eff}}{N} \right)^{1/2}. 
\label{eq:dtRN}
\ee
where $U^{\prime} \equiv dU/dt$. 
The second equality holds in 
the limit where noise decorrelates on  a time scale $\Delta$ 
much less than the pulse width, i.e. $\Delta \ll \fwhm$  for which
$\rho(t) =  \Delta \delta(t)$ and where the effective pulse width
is $W_{\rm eff} = 1 / \int\, dt [U^{\prime}(t)]^2$; the result is
written in terms of the SNR for a single pulse and the number of
pulses summed, $\SNR_1 = \SNR/\sqrt{N}$.
The third form shows
that the error does not depend on $\Delta$, where we have
used $S_{\rm sys}$ as the system equivalent flux density,
$S_{\rm peak}$ as the average peak flux density of a pulse,
and $B$ is the total bandwidth.   The factor of two accounts for
summing of two polarization channels.
 

For measured pulse profiles
$U_j = U(j\Delta t)$ sampled discretely at intervals 
$\Delta t \ll W_{\rm eff}$,
the effective width can be calculated by differencing,
\be
W_{\rm eff} = \frac{\Delta t} {\displaystyle \sum_j \left(U_{j+1}-U_j \right)^2}.
\label{eq:Weff}
\ee
The denominator is proportional to the structure function of
the pulse profile.
If the sampled pulse has finite signal-to-noise ratio, the denominator
must be corrected for the additive noise
 by subtracting $2n_{\rm on}\sigma_n^2$, where
$\sigma_n$ is the rms of the additive noise and $n_{\rm on}$ is the number of
terms in the sum. 
For a Gaussian pulse with width $\fwhm$ (FWHM), 
\be
\dtRN
= 
	\frac{\fwhm}{(2\pi\ln 2)^{1/4} \SNR_1 \sqrt N} 
	\left ( \frac{\Delta}{\fwhm}\right)^{1/2}.
\label{eq:radiometer}
\ee


We describe jitter by letting
$a(\phi)$ be the functional form of a single pulse and
$f_{\phi}(\phi)$ the probability density function (PDF) for the phase
pulse-to-pulse phase jitter.  
Assuming all pulses have the same shape, the ensemble-average
pulse shape is 
\be
U(\phi) \propto \int \, d\phi^{\prime} \, f_{\phi}(\phi^{\prime})
     a(\phi - \phi^{\prime}).
\ee
With this model and the phenomenologically consistent assumption
that an infinite pulse average converges to $U(\phi)$, 
the TOA error is (using the pulse period $P$) 
\be
\dtJ
= N_i^{-1/2} (1 + m_I^2)^{1/2} \,P \langle \phi^2 \rangle^{1/2}
     = N_i^{-1/2} (1 + m_I^2)^{1/2} \, P
       \left[ \int \, d\phi \, \phi^2 f_{\phi}(\phi)\right]^{1/2},
\ee
where $N_i\le N$ is the number of {\it statistically independent}
pulses that have been summed and $m_I$ is the 
intensity modulation index (ratio of rms intensity to mean intensity),
typically of order unity.
Intensity fluctuations without phase jitter yield no 
timing error but they enhance timing errors when there is phase jitter.
Letting $\langle \phi^2 \rangle^{1/2} \equiv \xi_U f_J W_i/P$ we define
a dimensionless jitter parameter 
$f_J$ equal to the width (FWHM)
of the phase-jitter PDF in units
of the resulting {\em intrinsic}  pulse width $\fwhm_i$,
where $\xi_U\sim 1$ is a shape-dependent factor.
For Gaussian shaped pulses with a Gaussian phase 
jitter distribution, we obtain
\be
\dtJ = \frac{f_J\fwhm_i(1+m_I^2)^{1/2}} {2 (2N_i \ln 2)^{1/2} }.
\label{eq:jitter}
\ee
This idealized result will also apply approximately to pulsars
with multicomponent profiles.  The amplitudes and phases of individual
subpulses in different components are only partially correlated,
so a profile with $M$ components would show a jitter-induced error
that is up to a factor $M^{-1/2}$ smaller than that for a single component.

\section{B. Scattering and Refraction from Thin Screens}
\label{app:screens}

Electron-density fluctuations on large scales refract, focus and
defocus radiation from a pulsar producing variations in flux density
and arrival time.   ``Large'' is defined relative
to the scale $\sim D\theta_d$ that is 
larger than the Fresnel scale $\sim\sqrt{\lambda D}$ by the same
factor that the Fresnel scale is larger than the diffraction scale
$\sim \lambda\thetad$. 
(e.g. Rickett 1990).    While an acceptable model for electron-density
variations is one with a continuum of length scales, it is useful to
separate large and small scales, as we do here.  The Kirchoff 
diffraction integral for the scalar wavefield 
has an integrand $\exp{[i\Phi(\xvec,\xvecp)]}$, where $\xvec$ and
$\xvecp$ are both two dimensional vectors transverse to the LOS.  
The total phase
$\Phi = \phi_g + \phi_r + \phi_d$ comprises
a geometric term
\be
\phi_g(\xvec,\xvecp) = \left (\frac{\pi\nu}{c} \right)
	\left[
	  \ds^{-1}{x^{\prime}}^2 + (D-\ds)^{-1} \vert \xvec - \xvecp\vert^2   
	\right],
\ee
a refraction term, $\phi_r$, from large-scale variations
 and the diffraction term, $\phi_d$, from small-scale variations.
$D$ is the Earth-pulsar distance and $\ds$ is the pulsar-screen distance. 
In the absence of the refraction term, the scattered wavefield of
a point source is described by an ensemble-average image
$I_0(\thetavec)$.  We include refraction as a quadratic surface centered
on a stationary-phase point $\xvecpbar$ on the screen 
corresponding to an angle $\thetavecbar$,
\be
\phi_r(\xvecp) =
	\phi_r(\xvecpbar)
	+ \bvec\cdot (\xvecp - \xvecpbar)
	+ (\xvecp - \xvecpbar) \cdot {\bf C} \cdot (\xvecp - \xvecpbar).
\label{eq:phir}
\ee 
The stationary phase point is defined with respect to 
$\phi_g + \phi_r$ and generally there could be multiple stationary phase
points. 
By consolidating linear and quadratic terms, the refraction-distorted image
can be written as
\be
I(\thetavec)  = I_0(
		G_x^{-1}(\theta_x-\thetabar_x),
	        G_y^{-1}(\theta_y-\thetabar_y)),
\ee
where the gains are given by
\be
G_{x,y} = \left[1 + k^{-1}(\ds/D)(D-\ds) C_{x,y} \right]^{-1}.
\ee
We assume, for simplicity, that ellipses of constant phase
from the last term in Eq.~\ref{eq:phir} align with the transverse
coordinate axes ($x,y$).  In this case, $\bf C$, is diagonal with
elements $C_{x,y}$.  A gain $G_{x,y} > 1$ implies that the image
is wider in the $x,y$ directions compared to the unity gain case. 
While the amplitude of the image is unaltered, the integral over the
image implies a flux variation by a factor $G_xG_y$. 
A more general expression that we use in Paper III is
\be
I(\thetavec) = I_0({\bf U}^{\dagger} \mbox{\boldmath $\gamma_2$}^{-1} {\bf U} 
	(\thetavec - \thetavecbar)), 
\ee
where ${\bf U}$ is a $2\times2$ matrix that diagonalizes ${\bf C}$,
the dagger denotes transpose, 
and $\gamma_2$ is a diagonal $2\times2$ matrix with elements
$G_x^{-1}$ and $G_y^{-1}$.

For each point on the screen the time delay relative to the direct ray
in vacuum is
\be
 t = t_{\DM}
	+ \frac{1}{2c}\left[\frac{D(D-\ds)}{\ds} \right]
	\left[
		  \thetabar^2 
		+ G_x^{-1}(\theta_x-\thetabar_x)^2 
		+ G_y^{-1}(\theta_y-\thetabar_y)^2 
	\right].
\ee
The first term $t_{\rm DM} = (2\pi\nu)^{-1} \phi_0(\xvecpbar)$ is the extra
dispersive delay from the position of the image centroid.   The second
term is the geometric delay from the centroid of the refracted image. 
The third term results from the combination of extra geometrical delay
and modified dispersive delay from the quadratic part of the refractive
phase.  A gain $G_x>1$ requires a converging wavefront that is produced
by a deficit of electrons compared to the local mean in the screen.  Thus
while $G_x > 1$ increases the size of the image in the $x$ direction
and increases geometric delays $\propto G_x^2$, 
these are compensated in part by
the reduced dispersion delay. 
  
We average the last term over the distorted image, which we
treat as a probability density function with angular variances
$(G_{x}{\sigma_{x}}_0)^2$ and $(G_{y} {\sigma_{y}}_0)^2$, 
yielding
\be
 t = t_{\DM}
	+ \frac{1}{2c}\left[\frac{D(D-\ds)}{\ds} \right]
	\left[
		  \thetabar^2 
		+ G_x {\sigma_x}_0^2
		+ G_y {\sigma_y}_0^2
	\right].
\ee
When the undistorted image is symmetric we have
 ${\sigma_x}_0 = {\sigma_y}_0$
and the time delay then involves a factor that is the
the sum of the gains, $G_x + G_y$. 

\section{C. Frequency Dependent Dispersion Measure from Phase-screen 
Averaging}
\label{app:cone}

Scattered pulsar radiation reaches the Earth from
a cone of radius $\theta_d$, the diffaction angle. 
For simplicity here we assume scattering occurs in a screen 
a distance $\Dp=D-\ds$ from the Earth.   The screen phase $\phi(\xvec)$
is averaged by the area of the scattering cone $\sim \ell_s^2$ 
where $\ell_s = \Dp\theta_d$. \DM\ is related to the screen phase by 
$\DM = \nu\phi/2\pi a_{\rm DM}$ where $a_{\DM}$ is defined
in Eq.~\ref{eq:tDM}.  The screen-averaged \DM\ is
\be
\DM(\nu) = \int d\xvec W(\xvec - \xvec_0, \nu) \DM(\xvec),
\ee 
where the averaging function $W(\xvec)$ has unit area and has a width
that is strongly frequency dependent.  Letting $\xvec_0$ be the
point on the screen where the direct ray path emerges with
an associated dispersion measure $\DM_0 = \DM(\xvec_0)$,
we derive the variance of the difference
$\delta \DM(\nu) = \DM(\nu) - \DM_0$.  Assuming that the screen
phase has stationary statistics,
that scattering is circularly symmetric, and assuming that 
the weighting function
has an autocorrelation function $R_W(\xvec, \xpvec, \nu)$ that depends
only on the difference $\xvec-\xpvec$, we can write the DM variations
in terms of the phase structure function $D_{\phi}(\delta\xvec)$
\be
\left\langle
[\delta\DM(\nu)]^2
\right\rangle
= \frac{1}{2\pi} \left(\frac{\nu}{a_{\DM}} \right)^2
\int dy\,y\, D_{\phi}(y)
\left[ W(y,\nu) - \frac{1}{2}R_W(y,\nu)\right].
\ee
If we use the Kolmogorov form $D_{\phi}(y)\propto y^{5/3}$
(see main text and \cite{cl91, cr98}) and if we approximate
the weighting function as $\ell_s^{-2}$ that drops to zero
at a radius $\ell_s$, we obtain
\be
\left\langle
[\delta\DM(\nu)]^2
\right\rangle 
= \frac{3f_{5/3}}{22\pi^2} \left(\frac{r_e c}{a_{\DM}} \right)^2
	\SM \ell_s^{5/3}.
\ee
Using $\ell_s = \Dp \theta_d$ and the relation
for a Kolmogorov medium in strong scattering
\citep{cl91, cl02} 
\be
\theta_d = 0.071~{\rm arc~sec}~ \SM^{3/5}\nu^{-11/5},
\ee
we obtain 
\be
\left\langle
[\delta\DM(\nu)]^2
\right\rangle^{1/2}   
\approx  10^{-1.05}~\DMu~\SM {\Dp}^{5/5}\nu^{-11/6}.
\ee
The associated rms time delay is
\be
\Delta t_{\delta\DM} = 4.15\times10^{-3}~s~\nu^{-2} 
	\left\langle [\delta\DM(\nu)]^2 \right\rangle^{1/2} 
	\approx
	0.118~\mu s 
	{\Dp}^{5/6} \nu^{-23/16}
	\left(\frac{\SM}{10^{-3.5}}~\SMu \right).
\ee

\section{D.  Least-squares Fitting of Multifrequency Data}
\label{app:lsq}

Consider a set of multi-frequency TOA measurements obtained in
an observing session at a given epoch,
$\left\{ t_k, \nu_k, k=1, n_{\nu} \right\}$.  
We model the TOAs as
$t(\nu) = t_{\infty} + \tDM(\nu) + t_C(\nu) + \epsilon(\nu)$ 
where $t_{\infty}$ is the TOA at infinite frequency,
$\tDM$ is the delay from plasma dispersion,
$t_C$ is an additional, non-dispersive chromatic perturbation
(e.g. from scattering), and 
$\epsilon$ is measurement error that we consider to be white noise
(statistically independent).  Actually, $t_C$ is the sum of several
effects as outlined in the main text, but for convenience here
we use only one term. 
In vector form the TOAs are  
$
\Dvec = \Xarray \thetavec + \epsvec,
$
where $\Dvec= {\rm col}\{ {\rm t_k} \}$ is a column vector of arrival times,
$\thetavec$ is a vector of parameters, and 
$\Xarray$ comprises the independent variables.   
With frequency-dependent errors described by 
a  covariance matrix $\Carray$, the solution is 
$\thetavec = 
	\left(\Xarray^{\dagger}\Carray^{-1}\Xarray\right)^{-1} 
	\Xarray^{\dagger}\Carray^{-1}\Dvec$.
The covariance matrix for parameter errors is
${\bf P} = \left(\Xarray^{\dagger}\Carray^{-1}\Xarray\right)^{-1}$
if the model is correct (only additive errors to the TOAs).
For white-noise errors in TOAs at the different 
frequencies, 
$\Carray = {\rm diagonal} \left\{ {\sigma_{\rm }}^2_k \right\}$.
In the following we will use the quantities
\be
r_p \equiv \sum_k \sigma_k^{-2} \nu_k^{-p}\ln\nu_k,
\quad\quad
s_p \equiv \sum_k \sigma_k^{-2} \nu_k^{-p}.
\label{eq:sp}
\ee
If we fit only for $t_{\infty}$ and there are no chromatic effects
(the situation at very high radio frequencies and
in high-energy observations) 
the variance
is $\sigma_{t_{\infty}}^2 = 1/s_0$.  When the errors are identical
at all frequencies, $\sigma_{t_{\infty}}^2 = \sigma^2/n_{\nu}$. 

{\bf Fit for dispersion:}  Consider a two-parameter fit
that includes the dispersion term along with $t_{\infty}$ 
and that there are no
other chromatic effects in the data. The parameter vector is 
$\thetavec = {\rm col} \left(t_{\infty}, a_{\DM} \DM \right)$,
the variables matrix is
${\bf X} = {\rm matrix}(1 \,\,\, \nu_i^{-2}), i=1,\ldots,n_{\nu}$,
and the standard errors are 
$\sigma_{t_{\infty}}^2 = s_4 / {\cal D}_2$  
for $t_{\infty}$ 
and
$\sigma_{\DM}^2 = s_0 / a_{\rm DM}^2 {\cal D}_2$ 
for \DM, where ${\cal D}_2 =  s_0 s_4 - s_2^2$
is the determinant of  
$\Xarray^{\dagger}\Carray^{-1}\Xarray$.

An alternative approach uses an ancillary value of DM
(e.g.  interpolated from measurements at other epochs)
that will have an error $\delta \DM$.
The TOA error is then
$
\sigma_{t_{\infty}}^2 = s_0^{-1} + 
		\left(a_{\DM}\delta\DM s_2/s_0\right)^2, 
$
which is better than the two-parameter fit if
$\delta\DM < 
	a_{\DM}^{-1} 
	\left(s_0 / {\cal D}_2\right)^{1/2}
$
Evaluating in the continuous limit 
with errors that are independent of frequency 
and letting $\nu_2 = \nu_1 - B$ where $B$ is the total bandwidth
and $\sigma_B = \sigma/\sqrt{n_{\nu}}$, we require
\be
\delta\DM < \frac{\sigma_B}{a_{\DM}}
	\frac{\sqrt{3}\nu_1^3}{B}(1-B/\nu_1)^{3/2}
	 =  10^{-3.38}\,\DMu\,
	\frac{\nu_1^3}{B}(1-B/\nu_1)^{3/2}
	\sigma_{B}(\mu s)
\ee
for $\nu_1$ and $B$ in GHz.

Now consider the case where the data contain the chromatic term
$t_C(\nu)$ but the fitting function does not. 
If $t_C(\nu)$ is deterministic\footnote{
Stochastic $t_C$ will yield an error in $t_{\infty}$ that involves
the correlation function $\langle t_C(\nu) t_C(\nup)\rangle$, but we will
not elaborate on this case in this paper.} 
 the additional, systematic errors are
\be
\delta t_{\infty} = 
		{\cal D}_2^{-1}
		\sum_k \sigma_k^{-2}(s_4-s_2\nu_k^{-2}) t_C(\nu_k)
		= -a_C R_{t_{\infty}} 
		= a_C {\cal D}_2^{-1} 
		  \left({s_4 s_X - s_2 s_{X+2}}\right)
\label{eq:dtinfty1}
\\
\delta \DM = 
		{\cal D}_2^{-1}
		\sum_k \sigma_k^{-2}(s_0\nu_k^{-2}-s_2) t_C(\nu_k)
		= 
		{a_C}\left(a_{\DM} {\cal D}_2\right)^{-1}
                \left(s_0 s_{X+2} - s_2 s_{X}\right)
\ee
where the last equality in each equation holds for a power law scaling,  
$t_C(\nu) = a_c\nu^{-X}$. 
Because it absorbs some of the scattering delay, $t_C$,
$\delta\DM$ is always positive, 
causing TOAs to be overcorrected and producing 
a negative bias in $t_{\infty}$. 

A simple case is where narrowband measurements are made at two
frequencies $\nu_{1,2}$ with $r=\nu_1/\nu_2$.  
The error from the chromatic term is
$
\delta t_{\infty} = 
	\left(r^2 t_{C_1} - t_{C_2} \right) / 
	\left(r^2-1\right).
$
For an octave bandwidth ($r=2$),   
the TOA error is approximately $-1/3$ of the chromatic term at the
lower frequency when $\nu_1 \gg \nu_2$ so that $t_{C_2} \gg t_{C_1}$.  
Not surprisingly,
observations over an octave at low frequencies where $t_C$
is larger will be substantially worse than at high frequencies.  
For large $r$ we have $\delta t_{\infty} \approx  - r^{X-2} t_{C_1}$, 
showing that the chromatic term at the higher frequency is
amplified by a large factor when $X=4$.

For $n_{\nu} = 1024$, $\nu_1 = 2$~GHz and $\nu_1 = 1$~GHz,  
we get $\delta t_{\infty} / a_c  = -0.29$
and $\delta\DM a_{\DM} / a_C = 1.15$.    Suppose that
$a_C = 1~\mu s$ if all frequencies are expressed in GHz,
equivalent to $t_C(1~{\rm GHz}) = 1~\mu s$.   In this case
$\delta t_{\infty} = -0.29~\mu s$ and
$\delta\DM = (1.15~\mu s / 4.15~{\rm ms})\DMu = 10^{-3.62}~\DMu$. 
It may be more convenient to express the systematic errors in terms
of $t_C$ at either the highest or lowest frequency, in which case
the relations $a_C = t_C\nu_1^X = t_C\nu_2^X$  can be used.

Various trends can be identified by considering other cases.
For fixed total bandwidth $B$, the systematic TOA error 
decreases with increasing upper frequency $\nu_1$.    
For fixed upper frequency, the {\em systematic error} increases as $B$
gets larger (i.e. as the lower frequency gets lower).  Counteracting
this trend is the decrease in {\em standard error} with larger bandwidth,
implying that there is an optimal bandwidth that depends on the
pulsar spectrum and system noise vs. frequency along with the scattering
strength. 

{\bf Fit for dispersion and scattering:} Now we include scattering
as a single term in the fitting function,
$t_C(\nu) = a_{\rm C} \nu^{-X}$, and assume initially
 that the exponent $X$ is known.
Then 
$\thetavec = {\rm col} \left(t_{\infty}, a_{\DM} \DM, a_{\rm C} \right)$ 
and the variables matrix is
${\bf X}={\rm matrix}(1 \,\,\, \nu_i^{-2}\,\, \nu_i^{-X}), i=1,\ldots,n_{\nu}$.


When $X$ is assumed to be $X_0$, the error  
$\delta X = X_0 - X$, 
causes a systematic error  from an extra term 
obtained from the expansion, 
$t_C(\nu; X_0) \approx 
	a_{\rm C} \nu^{-X} - a_{\rm C}  \delta X \nu^{-X} \ln\nu  $.
The error on the arrival time to linear order in $\delta X$ is
\be
\delta t_{\infty} = -a_C \delta X G_{t_{\infty}}
\ee
where
\be
G_{t_{\infty}}
=  {\cal D}_3^{-1}  
\left[
	{r_X}\, 
		\left(s_4 s_{2X} - s_{X+2}^2 \right) 
	{r_{X+2}}\, 
		\left(s_X s_{X+2} - s_2 s_{2X}  \right) 
	+ {r_{2X}}\, 
		\left(s_2 s_{X+2} - s_4 s_X \right) 
\right]
\ee
and 
\be
{\cal D}_3 = 
{\rm det~{\bf  \Xarray^{\dagger}\Carray^{-1}\Xarray}}
= (s_0s_4-s_2^2)s_{2X}
			+ 2 s_2 s_X s_{X+2}
			- (s_4 s_X^2+s_0s_{X+2}^2).
\ee

Evaluting for the same case as above with $n_{\nu} = 1024$,
$\nu_1 = 2$~GHz and $\nu_2 = 1$~GHz, we obtain
$G_{t_{\infty}} = -0.069$
implying for $\delta X > 0$ 
that, opposite to before, the TOA is underrcorrected because \DM\ 
is biased low while the scattering parameter $a_C$ is biased
high.  When $\delta X<0$ the opposite is true.
Either way, the amplitude of the timing error is small,
$\delta t_{\infty} = 0.069 a_C \delta X$.

Finally, a four parameter fit can include $\delta X$ in an iterative
approach that uses an initial value for $X$ and interates.
The parameter vector is
$
\thetavec = 
  {\rm col} \left(t_{\infty}, a_{\DM} \DM, a_{\rm C}, a_{\rm C_0}\delta X \right) 
$
and the variables matrix is
${\bf X}={\rm matrix}
(1 \,\,\, \nu_i^{-2}\,\, \nu_i^{-X}\,\, -a_{\rm C_0}\nu_i^{-X_0}\ln \nu_i), i=1,\ldots,n_{\nu}.$
In this case, initial choices for $a_{\rm C_0}$ and $X_0$ 
are updated by fitted values for $a_{\rm C}$ and $\delta X$.


\end{document}